\documentclass[10pt]{iopart}
\usepackage{bm,color}
\usepackage{graphicx}
\pdfoutput=1

\begin{document}
\title{Dynamical magnetoelectric phenomena of multiferroic skyrmions}
\author{Masahito Mochizuki$^{1,3}$, and Shinichiro Seki$^{2,3}$}
\address{
$^1$ Department of Physics and Mathematics, Aoyama Gakuin University, Kanagawa 252-5258, Japan\\
$^2$ RIKEN Center for Emergent Matter Science (CEMS), Wako 351-0198, Japan\\
$^3$ PRESTO, Japan Science and Technology Agency (JST), Tokyo 102-0075, Japan}
\ead{mochizuki@phys.aoyama.ac.jp}
\vspace{10pt}
\begin{indented}
\item[]March 2015
\end{indented}
\begin{abstract}
Magnetic skyrmions, vortex-like swirling spin textures characterized by a quantized topological invariant, realized in chiral-lattice magnets are currently attracting intense research interest. In particular, their dynamics under external fields is an issue of vital importance both for fundamental science and for technical application. Whereas observations of magnetic skyrmions had been limited to metallic magnets so far, their realization was discovered also in a chiral-lattice insulating magnet Cu$_2$OSeO$_3$ in 2012. Skyrmions in the insulator turned out to exhibit multiferroic nature with spin-induced ferroelectricity. Strong magnetoelectric coupling between noncollinear skyrmion spins and electric polarizations mediated by relativistic spin-orbit interaction enables us to drive motion and oscillation of magnetic skyrmions by application of electric fields instead of injection of electric currents. Insulating materials also provide an environment suitable for detection of pure spin dynamics through spectroscopic measurements owing to absence of appreciable charge excitations. In this article, we review recent theoretical and experimental studies on multiferroic properties and dynamical magnetoelectric phenomena of magnetic skyrmions in insulators. We argue that multiferroic skyrmions show unique coupled oscillation modes of magnetizations and polarizations, so-called electromagnon excitations, which are both magnetically and electrically active, and interference between the electric and magnetic activation processes leads to peculiar magnetoelectric effects in a microwave frequency regime.
\end{abstract}
%
\ioptwocol
\section{Introduction}
\label{sec1}
\subsection{Magnetic Skyrmions}
Noncollinear spin textures in magnets often host intriguing physical phenomena and useful device functions through coupling to charge degree of freedom, and thereby have been subject to intensive studies. For instance, it is known that magnetic domain walls and magnetic vortices can be driven by electric currents via a spin transfer torque mechanism, which points to their application for spin-electronics devices such as race track memory. Multiferroics with magnetically-induced ferroelectricity is another typical example~\cite{Fiebig05,Tokura06,Khomskii06,Cheong07,Tokura07,Tokura14}. Magnetic spirals in insulating magnets with frustrated interactions often induce asymmetry of charge distribution and resultant ferroelectric polarization $\bm P$ via relativistic spin-orbit interactions~\cite{Katsura05,Mostovoy06} in compounds such as TbMnO$_3$. The magnetoelectric coupling between spins and electric polarizations gives rise to rich cross-correlation phenomena such as magnetic-field switchings of ferroelectric polarization~\cite{Kimura03}, and electric-field controls of spin chiralities~\cite{Yamasaki07,Seki08}, magnetic modulation vectors~\cite{Murakawa09}, and magnetic domain distributions~\cite{Tokunaga09}.

\begin{figure}
\begin{center}
\includegraphics[width=1.0\columnwidth]{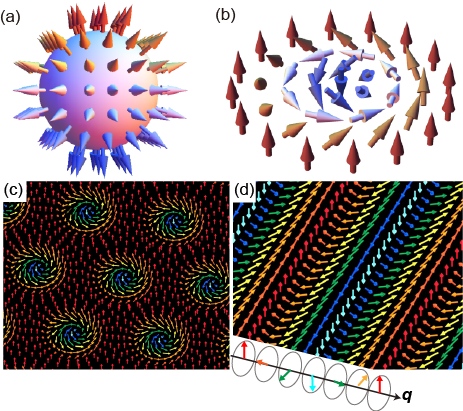}
\end{center}
\caption{(color online). (a) Schematic of the original hedgehog-type skyrmion. Its magnetizations point in all directions wrapping a sphere. (b) Schematic of the vortex-type skyrmion discovered in chiral-lattice ferromagnets, which corresponds to a projection of the hedgehog-type skyrmion onto a two-dimensional plane. Its magnetizations also point everywhere wrapping a sphere. (c) Schematic of a skyrmion crystal realized in chiral-lattice ferromagnets under an external magnetic field, in which skyrmions are hexagonally packed to form a triangular lattice. (d) Helical spin structure (or proper screw spin structure) with magnetizations rotating within a plane normal to propagation vector $\bm q$.}
\label{Fig01}
\end{figure}
In addition to these magnetic structures, magnetic skyrmions as nanometric spin whirls are recently attracting intensive research interest~\cite{Pfleiderer11,Nagaosa13,Fert13}. Skyrmion was originally proposed by Tony Skyrme in 1960s to account for stability of baryons in particle physics as a topological solution of a nonlinear sigma model in three dimensions~\cite{Skyrme61,Skyrme62}. In the original sense, a magnetic skyrmion comprises spins pointing in all directions wrapping a sphere similar to a hedgehog as shown in Fig.~\ref{Fig01}(a).

Later Bogdanov and his collaborators theoretically predicted that skyrmions should appear in ferromagnets without spatial inversion symmetry as vortex-like spin structures shown in Fig.~\ref{Fig01}(b), which corresponds to a projection of the original hedgehog skyrmion onto a two-dimensional plane~\cite{Bogdanov89,Bogdanov94,Rossler06}. They also predicted that skyrmions are often crystallized to form a triangular lattice as shown in Fig.~\ref{Fig01}(c), which is called skyrmion crystal. 

Magnetic interactions between two neighboring magnetizations in magnets are mostly categorized into two types, i.e., $\bm m_i \cdot \bm m_j$ type symmetric exchange interactions and $\bm m_i \times \bm m_j$ type Dzyaloshinskii-Moriya interactions. In magnets with broken inversion symmetry, there exists a finite net component of the Dzyaloshinskii-Moriya interaction~\cite{Dzyaloshinskii58,Moriya60}, which is given in a continuum from as,
\begin{eqnarray}
\mathcal{H}_{\rm DM} \propto \int \mathrm{d} \bm r 
\bm M \cdot (\bm \nabla \times \bm M),
\label{eqn:DMint}
\end{eqnarray}
where $\bm M$ is a classical magnetization vector. This interaction favors noncollinear alignment of magnetizations with 90$^\circ$ rotation with fixed handedness (spin chirality) and, thus, competes with the ferromagnetic exchange interaction which favors parallel (collinear) alignment of magnetizations. Consequently, a ground state of these chiral-lattice ferromagnets in the absence of external magnetic field is a helical state, so-called proper screw state, in which magnetizations rotate within a plane normal to the propagation vector $\bm q$ as shown in Fig.~\ref{Fig01}(d). An increase of the magnetic field at certain temperatures changes the helical phase to a skyrmion-crystal phase and eventually to a field-polarized ferromagnetic phase. A size of such skyrmions is typically 5-100 nm, which is determined by the ratio between strengths of the ferromagnetic interaction and the Dzyaloshinskii-Moriya interaction. The size becomes smaller for stronger Dzyaloshinskii-Moriya interaction.

A skyrmion is characterized by a topological invariant $Q$, so-called skyrmion number, which represents how many times the magnetizations wrap a sphere. The skyrmion number $Q$ is defined as,
\begin{eqnarray}
\int \mathrm{d}^2 r \left( 
\frac{\partial\hat{\bm n}}{\partial x} \times 
\frac{\partial\hat{\bm n}}{\partial y} \right)
\cdot \hat{\bm n}=\pm 4\pi Q,
\label{eqn:TopoInv}
\end{eqnarray}
where $\hat{\bm n}$ is a unit vector pointing in the local magnetization direction. The left-hand side of this equation represents a sum of solid angles spanned by three neighboring magnetizations. Because the magnetizations in a skyrmion point everywhere wrapping a sphere once, its value becomes $+4\pi$ or $-4\pi$. The sign is determined by the magnetization orientation at the core, that is, $Q=+1$ ($Q=-1$) for core magnetizations pointing upward (downward). 

This finite topological number indicates that skyrmions belong to a topological class distinct from non-topological classes, to which usual magnetic structures such as ferromagnetic states, helices and domain walls belong. This means that we cannot create a skyrmion starting from uniformly magnetized ferromagnetic state or annihilate a skyrmion through continuous modulation of the spatial magnetization alignment. Instead, a local flop of magnetization is required to create or annihilate a skyrmion, which costs rather large energy whose order is determined by the exchange interaction. Owing to this topological protection, skyrmion spin textures are very robust.

\subsection{Skyrmions in metallic B20 compounds}
\begin{figure}
\begin{center}
\includegraphics[width=1.0\columnwidth]{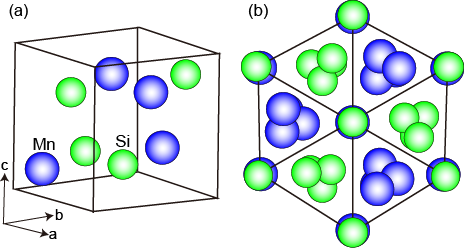}
\end{center}
\caption{(color online). (a) Crystal structure of MnSi with chiral cubic P2$_1$3 symmetry. (b) Chiral crystal structure of MnSi viewed along the [111] direction.}
\label{Fig02}
\end{figure}
In 2009, the realization of skyrmion crystals was discovered in a metallic B20 compound MnSi under an external magnetic field by the small-angle neutron scattering (SANS) measurements~\cite{Muhlbauer09}. Subsequently, real-space images of skyrmion spin structures and hexagonal skyrmion crystals were obtained by a Lorentz transmission electron microscopy (LTEM) for thin-film samples of Fe$_{1-x}$Co$_x$Si~\cite{YuXZ10}. The skyrmion-crystal phase was also observed in other B20 compounds such as FeGe and Mn$_{1-x}$Fe$_x$Ge by SANS~\cite{Pappas09,Pfleiderer10,Munzer10,Adams11,Grigoriev13} and LTEM~\cite{YuXZ11,Tonomura12,Shibata13,Morikawa13}. These compounds have a chiral crystal structure [see Figs.~\ref{Fig02}(a) and (b)], which belongs to the cubic P2$_1$3 space group. 
\begin{figure}
\begin{center}
\includegraphics[width=1.0\columnwidth]{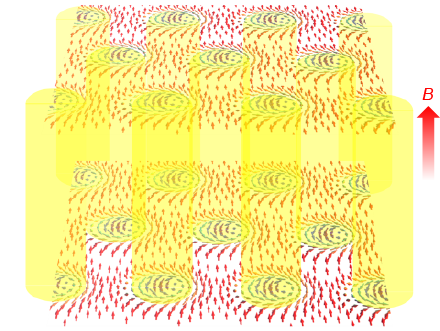}
\end{center}
\caption{(color online). Three-dimensional structure of magnetic skyrmion crystal where the skyrmion spin structures are stacked to form vortex tubes.}
\label{Fig03}
\end{figure}
In these materials, triangular skyrmion crystals appear on a plane normal to the applied magnetic field. Magnetizations in each skyrmion point antiparallel to the magnetic field at the center and gradually rotate upon propagating along the radial directions towards the periphery at which the magnetizations are parallel to the magnetic field. Such two-dimensional vortex-like textures are stacked to form tube-like structures as shown in Fig.~\ref{Fig03}.

\begin{figure}
\begin{center}
\includegraphics[width=1.0\columnwidth]{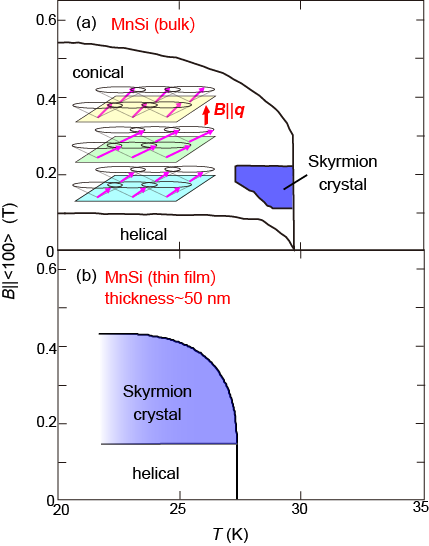}
\end{center}
\caption{(color online). Experimental phase diagrams of MnSi in plane of temperature $T$ and magnetic field $B$ for (a) bulk samples~\cite{Muhlbauer09} and (b) thin-film samples~\cite{Tonomura12}. For the bulk sample, the skyrmion-crystal phase appears only in a tiny region inside the conical phase at finite $T$ and $B$ on the verge of the boundary to the paramagnetic phase. In contrast, the skyrmion-crystal phase spreads over a wide $T$-$B$ range for thin-film samples. The enhanced stability of skyrmion-crystal phase is attributed to destabilization of the longitudinal conical phase in thin-film samples (see text). Inset in (a) shows the magnetic structure of the longitudinal conical state.}
\label{Fig04}
\end{figure}
Shown in Fig.~\ref{Fig04}(a) is the experimental phase diagram for bulk samples of MnSi in plane of temperature $T$ and magnetic field $B$~\cite{Muhlbauer09}. The skyrmion-crystal phase is positioned in a tiny window of $T$ and $B$ on the verge of the phase boundary between the paramagnetic and the helical (longitudinal conical) phases. This indicates that the skyrmion-crystal phase is rather unstable in the bulk samples. However the stability of the skyrmion-crystal state essentially depends on the dimension of the system, and it attains greater stability when the sample becomes thinner~\cite{YuXZ11,YiSD09,HanJH10,LiYQ11}. Figure~\ref{Fig04}(b) displays the experimental phase diagram for thin-film samples of MnSi~\cite{Tonomura12}. The area of the skyrmion-crystal phase noticeably spreads over a wide $T$-$B$ range, and is realized even at the lowest temperatures.

The enhanced stability of skyrmion crystal in thin-film samples can be understood as follows. When a magnetic field $\bm B$ is applied to a bulk sample, the conical spin structure propagating parallel to $\bm B$ with uniform magnetization component due to the spin canting towards the $\bm B$ direction is stabilized owing to energy gains from both the Dzyaloshinskii-Moriya interaction and the Zeeman coupling. The skyrmion-crystal state is usually higher in energy than this conical state. However, when the sample thickness becomes comparable to or thinner than the conical periodicity, the conical state can no longer benefit from the energy gain of the Dzyaloshinskii-Moriya interaction, and thus is destabilized. Instead the skyrmion-crystal state attains relative stability against the conical state. It has been argued that the uniaxial anisotropy, inhomogeneous chiral modulations, and the dipolar interaction can also stabilize skyrmions in thin-film samples~\cite{Butenko10,Kiselev11,Wilson12,Karhu12,Rybakov13,Kwon12}.

Skyrmions in chiral-lattice ferromagnets appear not only in the crystallized form but also as isolated defects in the ferromagnetic state~\cite{YuXZ10}. Such skyrmion defects are also stable because of the topological protection, and behave as particles since their spin textures are closed within a nanometre-scale region with peripheral magnetizations parallel to those of the outside ferromagnetic background. The isolated skyrmion defects are attracting a great deal of interest because they have turned out to possess numerous advantageous properties for technical application to information carriers for high-density and low-energy-consuming magnetic memories, that is, (1) topologically protected robustness, (2) small nanometric size, (3) rather high transition temperatures, and (4) ultra-low energy costs for driving their motion. Concerning the last property, it was experimentally revealed that one can drive their motion by injection of electric currents via the spin-transfer torques, and its threshold current density is five or six orders of magnitudes smaller than that for other noncollinear magnetic textures such as domain walls and helical structures~\cite{Jonietz10,YuXZ12,Everschor11,Everschor12}. Subsequent theoretical study attributed this high mobility or nearly pinning-free motion of skyrmions to their particle-like nature and finite topological number~\cite{Iwasaki13a,Iwasaki13b,Rosch13}. Skyrmions in metallic system have attracted research interest also for intriguing electron-transport phenomena due to the emergent electromagnetic field induced by their topological nature~\cite{Nagaosa12,LeeM07,Binz08,Neubauer09,ZangJ11,Kanazawa11,Schulz12,Milde13,Takashima14}. The skyrmion number directly corresponds to the gauge flux through the quantum Berry phase which gives rise to the topological Hall effect of conduction electrons.

\subsection{Skyrmions in magnetic insulator Cu$_2$OSeO$_3$}
While the observations of skyrmions in the early stage had been limited to specific metallic magnets with chiral B20 structure such as MnSi, FeGe, and Fe$_{1-x}$Co$_x$Si, formation of the skyrmion crystal was discovered also in an insulating magnet Cu$_2$OSeO$_3$ via the Lorentz transmission electron microscopy for thin-film samples~\cite{Seki12a} and the small angle neutron scattering experiments for bulk samples~\cite{Seki12b,Adams12}. 

\begin{figure}
\begin{center}
\includegraphics[width=1.0\columnwidth]{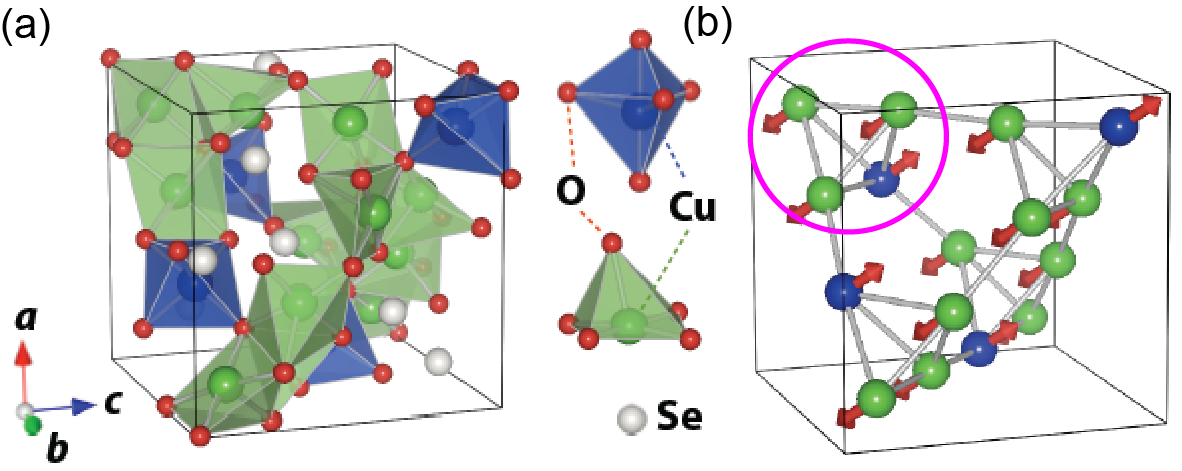}
\end{center}
\caption{(color online). (a) Crystal structure of the chiral-lattice magnetic insulator Cu$_2$OSeO$_3$, which belongs to the chiral cubic P2$_1$3 space group. (b) Magnetic structure of Cu$_2$OSeO$_3$ composed of tetrahedra of four Cu$^{2+}$ ions ($S$=1/2) with three-up and one-down spins. (Reproduced from Ref.~\cite{Seki12a}.)}
\label{Fig05}
\end{figure}
The crystal structure of Cu$_2$OSeO$_3$ belongs to the chiral cubic P2$_1$3 space group, which is equivalent to the B20 compounds [see Fig.~\ref{Fig05}(a)]. However the coordination of atoms in Cu$_2$OSeO$_3$ is quite different from that in the B20 compounds. There are two inequivalent Cu$^{2+}$ sites with different oxygen coordination. One is surrounded by a square pyramid of oxygen ligands, while the other is surrounded by a trigonal bipyramid. The nominal ratio between the former and the latter Cu$^{2+}$ ions is 3:1. The magnetic structure of Cu$_2$OSeO$_3$ consists of a network of tetrahedra composed of four Cu$^{2+}$ ($S$=1/2) ions at their apexes as shown in Fig.~\ref{Fig05}(b). Powder neutron diffraction~\cite{Bos08} and nuclear magnetic resonance~\cite{Belesi10,Belesi11} experiments suggested that three-up and one-down type ferrimagnetic spin arrangement is realized on each tetrahedron below $T_{\rm c}$$\sim$58 K.

\begin{figure}
\begin{center}
\includegraphics[width=1.0\columnwidth]{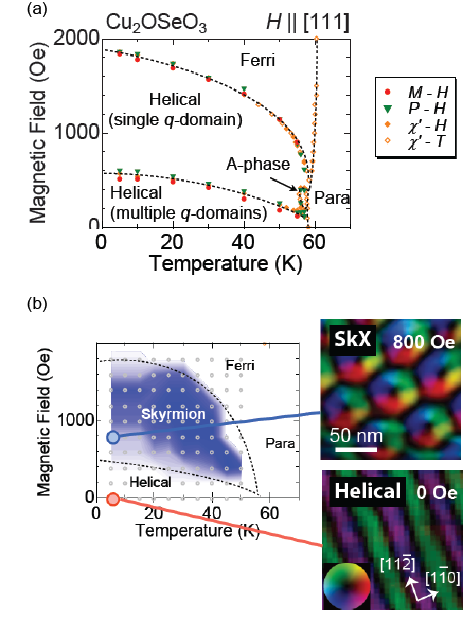}
\end{center}
\caption{(color online). Experimental $T$-$B$ phase diagrams of the copper oxoselenite Cu$_2$OSeO$_3$ for (a) bulk samples and (b) thin-film samples with sample thickness of $\sim$100 nm. In spite of the different origin of magnetism between metallic and insulating magnets, the phase diagrams of Cu$_2$OSeO$_3$ are similar to those of MnSi shown in Fig.~\ref{Fig04}. Whereas the skyrmion-crystal phase is restricted to a narrow $T$-$B$ window just below the magnetic-ordering temperature for the bulk samples, it spreads over the wide area and even to the lowest temperatures for the thin-film samples. Real-space images of the skyrmion crystal and the helical structure obtained by the Lorentz transmission electron microscopy are also displayed. (Reproduced from Ref.~\cite{Seki12a}.)}
\label{Fig06}
\end{figure}
Figures~\ref{Fig06}(a) and (b) show experimental $T$-$B$ phase diagrams for bulk samples and thin-film samples of Cu$_2$OSeO$_3$, respectively~\cite{Seki12a}. In spite of the different origin of magnetism between metallic and insulating magnets, the phase diagrams of the insulating copper oxoselenite Cu$_2$OSeO$_3$ are quite similar to those of the metallic B20 compounds. For bulk samples, the skyrmion-crystal phase takes place only as a small pocket (so-called A phase) in the phase diagram at finite $T$ and $B$. On the other hand, its area spreads over a wide $T$-$B$ range and even to the lowest temperatures for thin-film samples. Magnetic modulation periods in the helical and skyrmion-crystal states are $\sim 630$ \AA, which is much longer than the crystallographic lattice constant $\sim 8.9$ \AA.

\section{Magnetoelectric properties}
\label{sec2}
\subsection{Multiferroic nature}
In usual materials, their magnetic properties are affected or manipulated by magnetic fields, while their dielectric properties are by electric fields. In contrast, the control of magnetism by electric fields and inversely the control of dielectricity by magnetic fields are called magnetoelectric effect, which was first predicted by P. Curie more than a century ago~\cite{Curie1894}. The electric control of magnetism is one of the key issues in the field of spintronics aiming for low-energy-consuming magnetic devices because Joule-heating energy losses due to electric currents injected for generating a magnetic field or driving magnetic domains can be overcome by employing electric fields in insulators. The first experimental observation of the magnetoelectric effect was reported for Cr$_2$O$_3$ where the linear magnetoelectric effect ($M_i = \alpha_{ji} E_j$ and $P_i = \alpha_{ij} H_j$) shows up due to the simultaneous breaking of time-reversal and space-inversion symmetries but with a rather small coefficient $\alpha_{ij}$~\cite{Fiebig05,Tokura14}. In order to enhance the magnitude of the magnetoelectric effect, the employment of multiferroics, that is, materials endowed with both ferroelectric and magnetic orders, is particularly promising when these two orders are strongly coupled to each other. Recently several insulating magnets with noncollinear spin texture (such as TbMnO$_3$, MnWO$_4$, CuO, and a series of hexaferrites etc) have been reported to host ferroelectricity of magnetic origin~\cite{Kimura07}. In these materials, the low-symmetry of spin texture breaks spatial inversion symmetry inherent to the original crystal lattice, and causes a polar distribution of electric charges. The strong coupling between the spin texture and the ferroelectricity in such systems enables large and versatile magnetoelectric responses.

Whereas observations of magnetic skyrmion crystal had been limited to metallic B20 alloys, its realization was discovered also in a magnetic insulator Cu$_2$OSeO$_3$ (copper oxoselenite) with a chiral crystal symmetry in 2012. Noncollinear skyrmion spin textures in the insulator can induce ferroelectric polarization via relativistic spin-orbit interactions depending on the direction of external magnetic field. A presence or absence of the magnetism-induced ferroelectric polarization in the skyrmion-crystal phase and, if any, its direction can be known from the symmetry argument~\cite{Seki12a,Seki12c}. Although the spatial inversion symmetry is broken, the original crystal lattice of Cu$_2$OSeO$_3$ belongs to the non-polar space group P2$_1$3, and thus does not host ferroelectricity. In addition, the skyrmion-crystal spin structure itself as well as the conical and ferromagnetic (ferrimagnetic) spin orders are not polar, either. However combination of the crystal and the magnetic symmetries renders the system polar, and allows the emergence of ferroelectric polarization.

\begin{figure}
\begin{center}
\includegraphics[width=1.0\columnwidth]{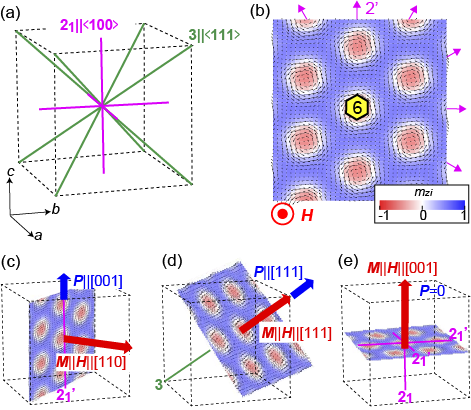}
\end{center}
\caption{(color online). (a) Symmetry axes of the Cu$_2$OSeO$_3$ crystal, which belongs to the chiral cubic P2$_1$3 space group: four three-fold rotation axes, 3, along $\left< 111 \right>$, and three two-fold screw axes, $2_1$, along $\left< 100 \right>$. (b) Magnetization configuration of the skyrmion crystal formed within a plane normal to the applied magnetic field $\bm H$, which possesses a six-fold rotation axis, 6, and six two-fold rotation axes followed by time reversal, $2'$. The in-plane magnetization components are represented by arrows, while the out-of-plane components are by colors. (c)-(e) When the skyrmion crystal is formed, most of the symmetry elements become broken, and the system can be polar to induce the ferroelectric polarization $\bm P$ depending on the direction of $\bm H$ (see text).}
\label{Fig07}
\end{figure}
As shown in Fig.~\ref{Fig07}(a), the crystal structure of Cu$_2$OSeO$_3$ possesses four three-fold rotation axes, 3, along $\left< 111 \right>$, and three $2_1$-screw axes along $\left< 100 \right>$. On the other hand, the magnetic structure of the skyrmion crystal has a six-fold rotation axis, 6, along the external magnetic field $\bm H$, and two-fold rotation axes followed by time reversal, $2'$, normal to $\bm H$ as shown in Fig.~\ref{Fig06}(b). Note that skyrmion crystals appear always on the plane normal to external magnetic field. When such a skyrmion spin texture is formed on the crystal lattice of Cu$_2$OSeO$_3$, most of the symmetry elements are broken, and eventually the system can become polar depending on the direction of $\bm H$.

We discuss three cases with different $\bm H$ directions shown in Figs.~\ref{Fig07}(c)-(e). When a skyrmion crystal sets in under $\bm H$$\parallel$[110] [see Fig.~\ref{Fig07}(c)], only the $2_1'$-axis ($\parallel$[001]) normal to $\bm H$ survives. Consequently the system becomes polar along [001], and emergence of ferroelectric polarization $\bm P$$\parallel$[001] is allowed. Likewise, in the case of $\bm H$$\parallel$[111] [see Fig.~\ref{Fig07}(d)], only the three-fold rotation axis parallel to $\bm H$ remains unbroken. Subsequently, emergence of $\bm P$$\parallel$$\bm H \parallel$[111] is allowed. In contrast, in the case of $\bm H$$\parallel$[001] as shown in Fig.~\ref{Fig07}(e), orthogonal arrangement of screw axes along $\langle 001 \rangle$ remains, and thus no ferroelectric polarization can be expected. This argument holds also for the helimagnetic and ferrimagnetic states, and one can expect emergence of ferroelectric polarizations $\bm P$$\parallel$[001] and $\bm P$$\parallel$[111] under $\bm H$$\parallel$[110] and $\bm H$$\parallel$[111], respectively, whereas $\bm P$ is zero under $\bm H$$\parallel$[001] for all these magnetic states. 

\begin{figure*}
\begin{center}
\includegraphics[width=2.0\columnwidth]{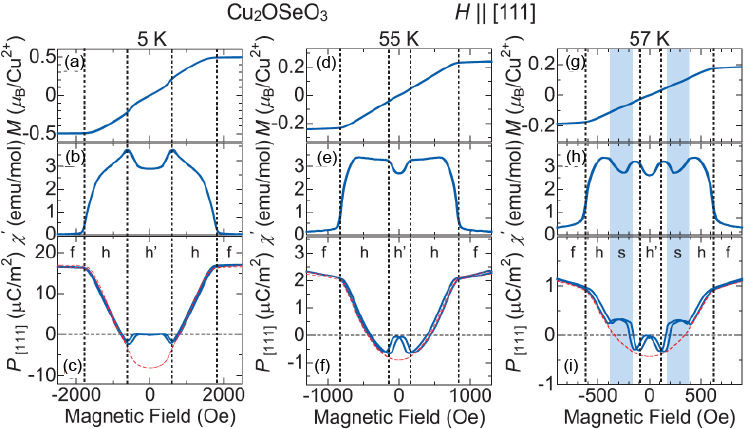}
\end{center}
\caption{(color online). (a)-(c) Magnetic-field dependence of (a) [111] component of net magnetization $M_{[111]}$, (b) ac magnetic susceptibility $\chi^\prime$, and (c) [111] component of ferroelectric polarization $P_{[111]}$ for bulk samples of Cu$_2$OSeO$_3$ measured under $\bm H$$\parallel$[111] at $T$=5 K. (d)-(f) Corresponding profiles at $T$=55 K. (g)-(i) Corresponding profiles at $T$=57 K. Letter symbols f, s, h, and h' stand for ferrimagnetic, skyrmion-crystal, helimagnetic (single $\bm q$-domain), and helimagnetic (multiple $\bm q$-domains) states, respectively. At zero magnetic field, the measured $P_{[111]}$ is zero because of cancellation of electric polarizations from multiple $\bm q$-domains. If one could obtain a single domain by field-cooling procedure, finite values of $P_{[111]}$ should be observed as indicated by dashed red lines. (Reproduced from Ref.~\cite{Seki12a}.)}
\label{Fig08}
\end{figure*}
\begin{figure*}
\begin{center}
\includegraphics[width=2.0\columnwidth]{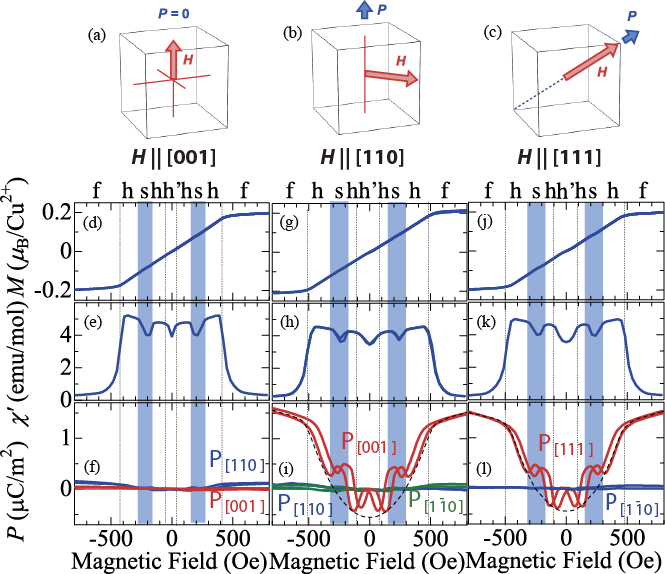}
\end{center}
\caption{(color online). (a)-(c) Magnetically-induced ferroelectric polarization $\bm P$ in Cu$_2$OSeO$_3$ for various directions of $\bm H$ predicted by the symmetry argument (see text). (d)-(f) Magnetic-field dependence of (a) net magnetization $M$, (b) ac magnetic susceptibility $\chi^\prime$, and (c) ferroelectric polarization $P$ for bulk samples of Cu$_2$OSeO$_3$ measured at 57 K for $\bm H$$\parallel$[001]. (g)-(i) Corresponding profiles for $\bm H$$\parallel$[110]. (j)-(l) Corresponding profiles for $\bm H$$\parallel$[111]. Letter symbols f, s, h, and h' stand for ferrimagnetic, skyrmion-crystal, helimagnetic (single $q$-domain), and helimagnetic (multiple $q$-domains) states, respectively. The measured $P$ is always zero at $H$=0 because of cancellation of contributions from multiple $\bm q$-domains. Finite values of $P_{[001]}$ and $P_{[111]}$ indicated by dashed lines should be observed in a single-domain sample after field-cooling procedure. (Reproduced from Ref.~\cite{Seki12c}.)}
\label{Fig09}
\end{figure*}
Figures~\ref{Fig08}(a)-(i) show $H$-dependence of [111] component of net magnetization $M_{[111]}$, ac magnetic susceptibility $\chi^\prime$, and [111] component of ferroelectric polarization $P_{[111]}$ for $\bm H$$\parallel$[111] measured at different temperatures, that is, 5 K, 55 K and 57 K~\cite{Seki12a}. The system goes through several magnetic phases as the magnetic field increases. At $T$=5 K and 55 K, the helimagnetic state (multiple $q$-domain), helimagnetic state (single $q$-domain), and collinear ferrimagnetic state successively emerge with increasing magnetic field. On the other hand, the phase evolution at $T$=57 K with increasing magnetic field is as follows: helimagnetic state (multiple $q$-domain) $\rightarrow$ helimagnetic state (single $q$-domain) $\rightarrow$ skyrmion-crystal state $\rightarrow$ helimagnetic state (single $q$-domain) $\rightarrow$ ferrimagnetic state. Namely the skyrmion-crystal phase takes place within the helimagnetic phase at $T$=57 K. The profile of $\chi^\prime$ shows clear anomalies at the magnetic transition points. Note that these magnetic-phase evolutions are not affected by the $\bm H$-direction. We find that all the magnetically ordered states can induce finite ferroelectric polarization $\bm P$ along [111] direction under $\bm H$$\parallel$[111] in agreement with the above symmetry analysis, but they have different signs and magnitudes. 

Figures~\ref{Fig09}(d)-(l) show $H$-dependence of net magnetization $\bm M$, ac magnetic susceptibility $\chi^\prime$, and ferroelectric polarization $\bm P$ measured at 57 K (just below $T_c \sim 58$ K) for different $\bm H$-directions, that is, $\bm H$$\parallel$[001], $\bm H$$\parallel$[110], and $\bm H$$\parallel$[111]~\cite{Seki12c}. A presence or absence of ferroelectric polarization $\bm P$ and its orientation for each $\bm H$ direction are totally consistent with the above symmetry argument [see Figs.~\ref{Fig09}(a)-(c)]. The phase evolutions and magnetoelectric nature of Cu$_2$OSeO$_3$ discussed here have been confirmed by several experimental techniques such as electron spin resonance (ESR) measurements~\cite{Maisuradze12}, magnetoelectric susceptibility measurements~\cite{Omrani14,Ruff15}, resonant soft x-ray scatterings~\cite{Langner14}, and muon-spin rotation measurements~\cite{Lancaster15}.

\subsection{Spin-dependent metal-ligand hybridization mechanism}
\begin{figure}
\begin{center}
\includegraphics[width=1.0\columnwidth]{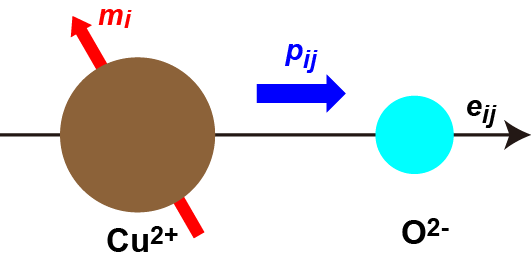}
\end{center}
\caption{(color online). Schematics of the spin-dependent metal-ligand hybridization mechanism as an origin of magnetism-induced electric polarizations. Local electric polarizations $p_{ij}$ emerge along the bond vector $\bm e_{ij}$ connecting the $i$th magnetic metal ion Cu$^{2+}$ and the $j$th ligand ion O$^{2-}$ whose magnitude depends on the relative direction of magnetization $\bm m_i$ against the bond.}
\label{Fig10}
\end{figure}
A theoretical study based on the first-principles calculation suggested a crucial role of relativistic spin-orbit interactions for magnetoelectric coupling in Cu$_2$OSeO$_3$~\cite{Yang12}. The electric polarizations in Cu$_2$OSeO$_3$ are microscopically induced via the so-called spin-dependent metal-ligand hybridization mechanism. Local electric polarizations $\bm p_{ij}$ induced by this mechanism are given by~\cite{Jia06,Jia07,Arima07},
\begin{eqnarray}
\bm p_{ij} \propto (\bm e_{ij} \cdot \bm m_i)^2 \bm e_{ij}.
\label{eq:sdepP1}
\end{eqnarray}
Here $i$ and $j$ are indices of the transition-metal ions Cu$^{2+}$ and the oxygen O$^{2-}$ ions, respectively. This mechanism assumes a single pair of adjacent magnetic (Cu$^{2+}$) and ligand (O$^{2-}$) ions with $\bm e_{ij}$ and $\bm m_i$ being the unit vector along the bond connecting them and the magnetization direction at the transition-metal site, respectively. The covalency between these two sites is governed by the relative direction of $\bm m_i$ against the bond through the spin-orbit interaction, and the local polarization $\bm p_{ij}$ is induced along the bond direction $\bm e_{ij}$.

\begin{figure}
\begin{center}
\includegraphics[width=1.0\columnwidth]{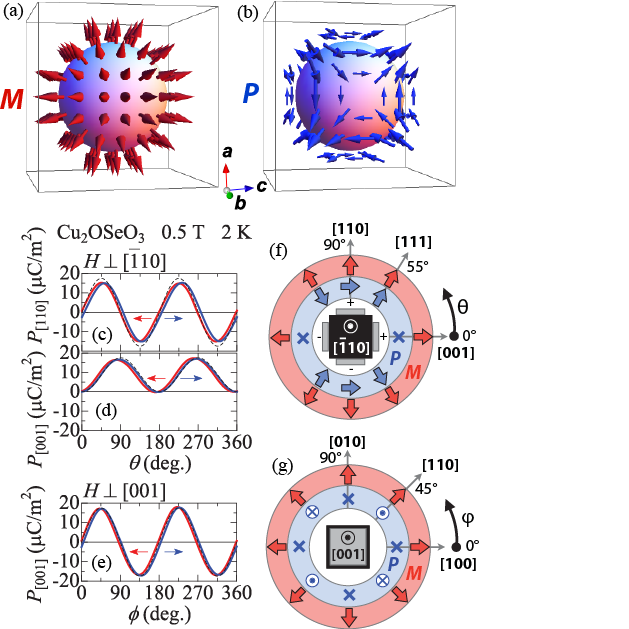}
\end{center}
\caption{(color online). (a), (b) Three-dimensional representation of general correspondence between (a) $\bm M$- and (b) $\bm P$-directions in the collinear spin state where arrows at the same position in (a) and (b) represent the $\bm M$-vector and corresponding induced $\bm P$-vector, respectively.
(c) [110] and (d) [001] components of ferroelectric polarization $\bm P$ simultaneously measured under a magnetic field $\bm H$ rotating around the [$\bar{1}$10] axis. (e) [001] component of $\bm P$ under $\bm H$ rotating around the [001] axis. Both measurements are performed for the collinear ferrimagnetic state at 2 K with $H$=0.5 T. Dashed lines indicate the theoretically expected behaviors from Eq.~(\ref{eq:sdepP1}), and arrows denote the direction of $\bm H$-rotation. (f),(g) Experimentally obtained relationships between the directions of $\bm P$ and $\bm M$ in the ferrimagnetic state and definitions of $\theta$ and $\phi$ (the angle between the $\bm H$-direction and the specific crystal axis) are summarized for (f) $\bm H$ rotating around [$\bar{1}$10] axis and (g) $\bm H$ rotating around [001] axis. Here the directions of $\bm M$ and $\bm P$ are indicated by thick arrows, while the cross symbol $\times$ denotes $P=0$. (Reproduced from Ref.~\cite{Seki12c}.)}
\label{Fig11}
\end{figure}
By taking a summation of contributions $\bm p_{ij}$ given by Eq.~(\ref{eq:sdepP1}) for all Cu-O bonds within a crystallographic unit cell, the local polarization $\bm p(\bm r)$ from the crystallographic unit cell can be evaluated. Since the modulation period of the magnetic skyrmion lattice is much longer than the crystallographic lattice constant in Cu$_2$OSeO$_3$, the local spin structure within a crystallographic unit cell can be regarded as nearly collinear. Consequently one-by-one correspondence between the local magnetization $\bm m(\bm r)$ and the local electric polarization $\bm p(\bm r)$ can be obtained as shown in Figs.~\ref{Fig11}(a) and (b)~\cite{Seki12c}.

For the collinear ferrimagnetic state with spatially uniform $\bm m(\bm r)$ and $\bm p(\bm r)$, one can know the dependence of ferroelectric polarization $\bm P$ on the $\bm H$-direction. Shown in Figs.~\ref{Fig11}(c) and (d) are the $\bm H$-direction dependence of the [110] and [001] components of $\bm P$ ($P_{[110]}$ and $P_{[001]}$) measured at 2 K with $H=0.5$ T, namely, in the ferrimagnetic phase. Here $\bm H$ rotates around the $[\bar{1}10]$-axis, and $\theta$ is defined as an angle between $\bm H$-direction and the [001] axis [see Fig.~\ref{Fig11}(f)]. The development of $P_{[001]}$ measured for $\bm H$ rotating around the [001] axis is also shown in Fig.~\ref{Fig11}(e) where an angle between $\bm H$ and the [100] axis is defined as $\phi$ [see Fig.~\ref{Fig11}(g)]. Both of the $\bm P$-profiles show sinusoidal modulation with a period of $180^\circ$ as a function of respective $\bm H$-rotation angle. On the other hand, the calculation using Eq.~(\ref{eq:sdepP1}) predicts $P_{[110]} \propto \sin 2\theta$ and $P_{[001]} \propto 1-\cos 2\theta$ for the former case, while $P_{[001]} \propto \sin 2\phi$ for the latter case. These behaviors are indicated by dashed lines in Figs.~\ref{Fig11}(c)-(e), which perfectly reproduce the experimentally observed  $\bm P$ profiles and hence strongly suggests a validity of the spin-dependent metal-ligand hybridization mechanism as an  origin of the electric polarizations in Cu$_2$OSeO$_3$.

\begin{figure}
\begin{center}
\includegraphics[width=1.0\columnwidth]{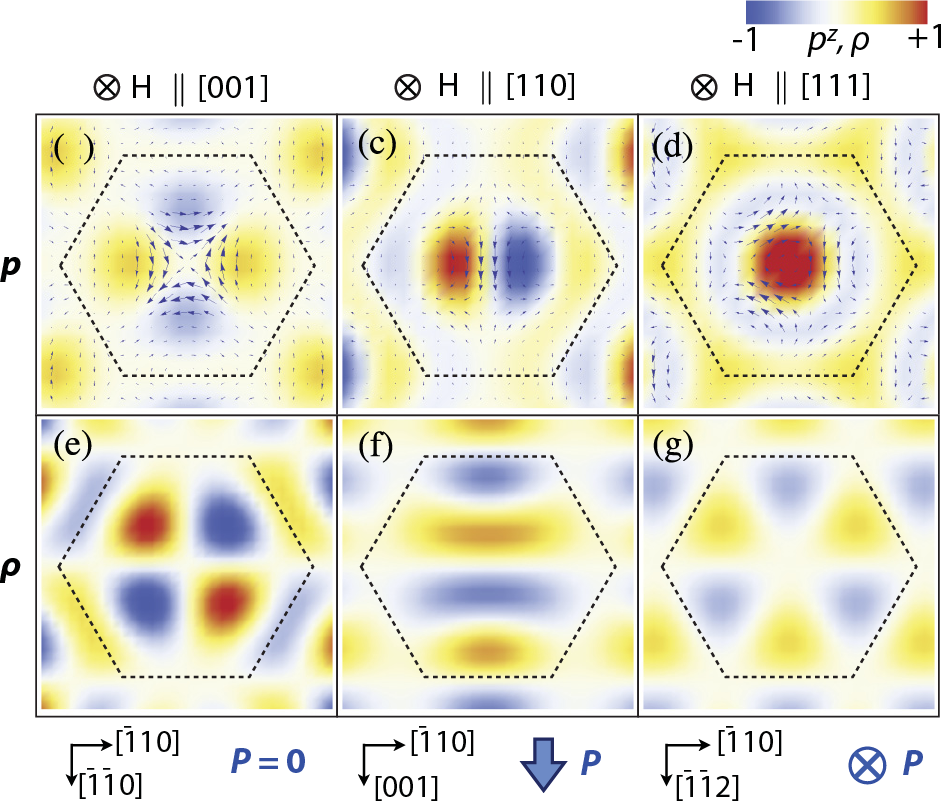}
\end{center}
\caption{(color online). Calculated spatial distributions of (a)-(c) local electric polarization vectors $\bm p$, and (d)-(f) local electric charges $\rho$ for the skyrmion-crystal state (see text). Magnetic field $\bm H$ is applied along the out-of-plane direction. They are for $\bm H$$\parallel$[001] [(a),(d)], $\bm H$$\parallel$[110] [(b),(e)], and $\bm H$$\parallel$[111] [(c),(f)]. The background color represents relative values of $p_z$ for (a)-(c) and $\rho$ for (d)-(f), respectively. Here $m_z$ and $p_z$ stand for the out-of-plane components of $\bm m$ and $\bm p$, respectively. The dashed hexagon indicates a magnetic unit cell of the skyrmion crystal or a single skyrmion. (Reproduced from Ref.~\cite{Seki12c}.)}
\label{Fig12}
\end{figure}
By identifying the microscopic magnetoelectric coupling given by Eq.~(\ref{eq:sdepP1}), spatial distributions of electric polarizations $\bm p(\bm r)$ and electric charges $\rho(\bm r)=-\bm \nabla \cdot \bm p(\bm r)$ can be obtained for a single skyrmion. The spatial distribution of $\bm m(\bm r)$ in the skyrmion crystal is approximately given by,
\begin{equation}
\bm m(\bm r) \propto \bm e_z M_0  + \sum^3_{n=1} 
[\bm e_z \cos(\bm q_n \cdot \bm r + \pi)
+\bm e_n \sin(\bm q_n \cdot \bm r + \pi)],
\nonumber \\
\label{EqSkyrmion}
\end{equation}
where $\bm q_n$ denotes three magnetic modulation vectors normal to $\bm H$ with relative angles of 120$^\circ$, and $\bm e_n$ is a unit vector orthogonal to $\bm e_z$ and $\bm q_n$ defined such that all $\bm q_n \cdot (\bm e_z \times \bm e_n$) have the same sign. Here $M_0$ scales with relative magnitude of the net magnetization in the $\bm H$-direction. Figure~\ref{Fig12} indicates real-space distributions of $\bm p(\bm r)$ and $\rho(\bm r)$ calculated for the skyrmion-crystal state with various directions of $\bm H$. The obtained results suggest that each skyrmion texture locally carries an electric quadrupole moment under $\bm H$$\parallel$[001] or an electric dipole moment along the in-plane ($\parallel$[001]) and out-of-plane ($\parallel$[111]) directions under $\bm H$$\parallel$[110] and $\bm H$$\parallel$[111], respectively. Such a local coupling between the electric dipole and the skyrmion spin texture strongly suggests that skyrmions in an insulator can be driven by a spatial gradient of the external electric field. Note that the total charge within each skyrmion is always zero, which implies nondissipative nature of the electric-field-induced skyrmion dynamics.

A small angle neutron scattering experiment under an applied electric field for bulk sample of Cu$_2$OSeO$_3$ indeed reported that orientation of $\bm q$ vectors in the skyrmion-crystal state slightly rotates around the $\bm H$-direction in a clockwise or counterclockwise manner depending on the sign of electric field~\cite{White12,White14}. This experiment clearly demonstrates possible manipulation of magnetic skyrmions by applying electric fields in insulators.

\subsection{Magnetoelectric coupling}
Because of the cubic crystal symmetry, the local electric polarization $\bm p_i$ for the $i$th crystallographic unit cell is described using the local magnetization components $\bm m_i = (m_{ia}, m_{ib}, m_{ic})$ as
\begin{eqnarray}
\bm p_i=
\left(
\begin{array}{c}
p_{ia}\\
p_{ib}\\
p_{ic}
\end{array}
\right)
= \lambda
\left(
\begin{array}{c}
m_{ib}m_{ic}\\
m_{ic}m_{ia}\\
m_{ia}m_{ib}\\
\end{array}
\right),
\label{eq:sdepP2}
\end{eqnarray}
where $a$, $b$ and $c$ are the Cartesian coordinates in the cubic setting~\cite{Mochizuki13,Liu13}. Although the two equations, Eq.~(\ref{eq:sdepP1}) and Eq.~(\ref{eq:sdepP2}), seem to be different from each other at first glance, it was confirmed that these two expressions give equivalent spatial distribution of the local polarizations. The magnetoelectric-coupling constant $\lambda$ is a material parameter, which is microscopically related with the relativistic spin-orbit interaction and the metal-ligand hybridization. Its value is common for all the magnetic phases as far as the same compound is concerned. Namely the value is identical in the helical, the skyrmion-crystal and the ferrimagnetic phases in Cu$_2$OSeO$_3$, and can be evaluated as $\lambda$=$5.64\times10^{-27}$ $\mu$Cm from the experimentally measured $P_{[001]}$=16 $\mu$C/m$^2$ in the ferrimagnetic phase under $\bm H$$\parallel$[110] at 5 K. The reason why we choose the ferrimagnetic phase is that all the tetrahedra with three-up and one-down spins give uniform contributions to the ferroelectric polarization $\bm P$ and the net magnetization $\bm M$. Thus the contributions from each tetrahedron, $\bm p_i$ and $\bm m_i$, can be easily evaluated from the experimentally measured $\bm P$ and $\bm M$ as $\bm p_i$=$\bm P$/$N$ and $\bm m_i$=$\bm M$/$N$ where $N$ is the number of the tetrahedra in the unit volume. In this way, the ferrimagnetic phase provides us a unique opportunity to evaluate the coupling constant $\lambda$.

\section{Spin model and phase diagrams}
\label{sec3}
\subsection{Microscopic spin model}
In 1980, Bak and Jensen proposed that magnetism on the chiral cubic crystal structure can be described by the following continuum spin model as long as magnetic orders with sufficiently slow spatial and temporal variations are concerned~\cite{Bak80}:
\begin{eqnarray}
\mathcal{H}=\int d\bm r & &\left[\frac{J}{2a}(\nabla \bm m)^2 
+\frac{D}{a^2}\bm m \cdot (\nabla \times\ \bm m) \right.
\nonumber \\
& &-\frac{g\mu_{\rm B}\mu_0}{a^3}\bm H \cdot \bm m 
\nonumber \\
& &+\frac{A_1}{a^3}(m_x^4 + m_y^4 + m_z^4) 
\nonumber \\
& &-\left. 
\frac{A_2}{2a}[(\nabla_x m_x)^2+(\nabla_y m_y)^2+(\nabla_z m_z)^2]
\right].
\nonumber \\
\label{Hcontinum}
\end{eqnarray}
This model describes competitions among the ferromagnetic exchange interaction (the first term), the Dzyaloshinskii-Moriya interaction (the second term), and the Zeeman coupling to an external magnetic field $\bm H$ (the third term). Two types of magnetic anisotropies allowed by the cubic crystal symmetry (the fourth and the fifth terms) are also incorporated, but they turn out to play only a minor role if we consider realistic small values of $A_1$ and $A_2$. In order to treat this continuum spin model numerically, it is convenient to divide the space into cubic meshes, which gives a classical Heisenberg model on the cubic lattice~\cite{YiSD09,HanJH10,Buhrandt13}. The Hamiltonian is given by,
\begin{eqnarray}
\mathcal{H}=
& &-J \sum_{i,\hat{\bm \gamma}} \bm m_i \cdot \bm m_{i+\hat{\bm \gamma}}
\nonumber \\
& &-D \sum_{i,\hat{\bm \gamma}} \left(
\bm m_i \times \bm m_{i+\hat{\bm \gamma}} \cdot \hat{\bm \gamma}
\right)
\nonumber \\
& &-g\mu_{\rm B}\mu_0 \bm H \cdot \sum_i \bm m_i
\nonumber \\
& &+A_1 \sum_{i} [(m_{i}^x)^4+(m_{i}^y)^4+(m_{i}^z)^4]
\nonumber \\
& &-A_2 \sum_{i} (m_{i}^xm_{i+\hat{x}}^x+m_{i}^ym_{i+\hat{y}}^y+m_{i}^zm_{i+\hat{z}}^z).
\label{eqn:model}
\end{eqnarray}
The index $\hat{\bm \gamma}$ runs over $\hat{\bm x}$, $\hat{\bm y}$, and $\hat{\bm z}$ for the three-dimensional case, while over $\hat{\bm x}$ and $\hat{\bm y}$ for the two-dimensional case. The magnetic system of Cu$_2$OSeO$_3$ is a network of tetrahedra composed of four Cu$^{2+}$ ($S$=1/2) ions, on which three-up and one-down type collinear spin arrangement is realized below $T_{\rm c}$$\sim$58 K~\cite{Bos08,Belesi10,Belesi11}. This four-spin assembly as a magnetic unit can be treated as a classical magnetization vector $\bm m_i$ whose norm $m$ is unity. We choose the ratio $D/J$=0.09, which gives a skyrmion diameter of $\sim$99 sites for the skyrmion-crystal phase. If we assume that the distance between adjacent tetrahedra is $\sim$5 \AA, this number corresponds to the skyrmion diameter of $\sim$50 nm in agreement with the observation in the Lorentz transmission electron microscopy for Cu$_2$OSeO$_3$~\cite{Seki12a}. For the slowly varying spin textures, the spins are nearly decoupled from the background lattice structure. It justifies the theoretical treatment based on the spin model on the cubic lattice for simplicity without considering the complicated crystal structure of real materials.

\subsection{Theoretical phase diagrams}
Although microscopic studies based on the first-principles calculations~\cite{Yang12,Janson14, Romhanyi14,Chizhikov15} and electron spin resonance (ESR) experiments~\cite{Ozerov14} have predicted complicated magnetic system with significant quantum nature, the above-introduced simple classical spin model has turned out to describe magnetic properties of Cu$_2$OSeO$_3$ well.
\begin{figure*}
\begin{center}
\includegraphics[width=2.0\columnwidth]{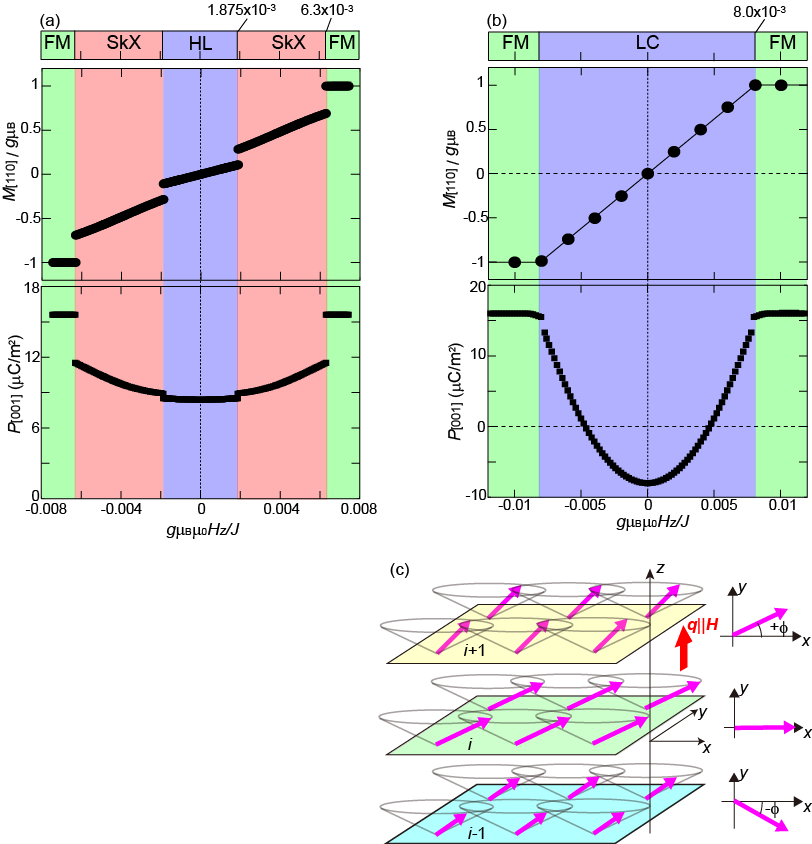}
\end{center}
\caption{(color online). (a) Theoretical phase diagram of the lattice spin model given by Eq.~(\ref{eqn:model}) with $D/J$=0.09 and $A_1$=$A_2$=0 for two dimensions as a function of magnetic field $H_z$ at $T$=0. Calculated magnetic-field dependence of [110] component of the net magnetization $M_{[110]}$ and that of [001] component of the ferroelectric polarization $P_{[001]}$ under $\bm H=(0,0,H_z)$($\parallel$[110]) are plotted. (b) Those for three dimensions. In the middle panel, $M_{[110]}$ data for the ground-state magnetic configurations obtained by numerically minimizing the energy are shown by closed circles, while the analytically obtained behavior of $M_z$=$\cos \theta$ with Eq.~(\ref{eq:theta}) and Eq.~(\ref{eq:cosphi}) is shown by a solid line. They show perfect coincidence. (c) Schematic illustration of the longitudinal conical spin structure.}
\label{Fig13}
\end{figure*}
Ground-state phase diagrams of the lattice spin model given by Eq.~(\ref{eqn:model}) well reproduce the experimental phase diagrams of Cu$_2$OSeO$_3$ as well as the B20 compounds. Figure~\ref{Fig13}(a) displays a theoretical phase diagram for two dimensions. We find that the skyrmion-crystal phase emerges in the range $1.875\times10^{-3}<|g\mu_{\rm B}\mu_0 H_z/J|<6.3\times10^{-3}$ sandwiched by the helical and ferromagnetic phases. This phase evolution is in agreement with an experimental result for thin-film samples shown in Fig.~\ref{Fig06}(b) where the helical, skyrmion-crystal, and ferrimagnetic phases successively appear as a magnetic field increases at low temperatures. For the relevant exchange parameter $J$=3 meV for Cu$_2$OSeO$_3$, these critical fields, respectively, correspond to 486 Oe and 1632 Oe, which coincide well with the experimental values of 500 Oe and 1800 Oe indicated by the phase diagram shown in Fig.~\ref{Fig06}.

The net magnetization $\bm M$ and the ferroelectric polarization $\bm P$ are given by sums of the local contributions as,
\begin{eqnarray}
\bm M=\frac{g\mu_{\rm B}}{NV}\sum_{i=1}^{N} \bm m_i,
\label{eq:netM}
\end{eqnarray}
and 
\begin{eqnarray}
\bm P=\frac{1}{NV}\sum_{i=1}^{N} \bm p_i.
\label{eq:netM}
\end{eqnarray}
Here the index $i$ runs over the Cu-ion tetrahedra with four spin pair, $N$ is the number of the tetrahedra, and $V$(=1.76$\times$10$^{-28}$ m$^3$) is the volume per tetrahedron. The calculated [110] component of $\bm M$ and the [001] component of $\bm P$ under $\bm H$$\parallel$[110] are plotted in the lower panels as functions of $H$.

It might seem to be surprising that the purely two-dimensional spin model can reproduce the experimentally observed phase evolutions in thin slab of materials with finite thickness. This is probably because the two-dimensional model captures well a physical mechanism for enhanced stability of skyrmion crystal due to the destabilization of conical state in thin-film samples as discussed in Sec.~1.2. However, because phase transitions take place only at zero temperature for the two-dimensional model, consideration of the three-dimensionality is required when we study physical properties of thin-film samples at finite temperatures.

On the other hand, a ground-state phase diagram of the spin model given by Eq.~(\ref{eqn:model}) for three dimensions is displayed in Fig.~\ref{Fig13}(b). In this case, the skyrmion-crystal phase is absent at low temperatures, and only a phase transition from the longitudinal conical phase to the ferromagnetic phase is reproduced at $|g\mu_{\rm B}\mu_0 H_z/J|=8.0\times10^{-3}$. This result again agrees with the experimental result for bulk samples shown in Fig.~\ref{Fig06}(a) where the skyrmion-crystal phase does not appear at low temperatures. 

The critical magnetic field $|g\mu_{\rm B}\mu_0 H_z/J|=8.0\times10^{-3}$ corresponds to 2073 Oe if we assume $J$=3 meV and $m$=1, which is again in good agreement with the experimental value of 1900 Oe. Note that there appears a multiple $\bm q$ conical phase in the experimental phase diagram at low-field region owing to degeneracy of the spiral $\bm q$ vectors and higher-order magnetic anisotropies, whereas it is absent in the present theoretical phase diagram because we neglect magnetic anisotropies for simplicity. It should also be mentioned that a recent Monte-Carlo study demonstrated that this three-dimensional lattice spin model can reproduce whole experimental phase diagram for bulk samples of Cu$_2$OSeO$_3$ and MnSi in plane of magnetic field $B$ and temperature $T$ with small skyrmion-crystal phase near the paramagnetic--conical phase boundary at finite temperatures~\cite{Buhrandt13}.

In the longitudinal conical state, the ferromagnetically ordered planes are stacked along the $\bm H$ direction accompanied by rotation of the in-plane magnetization components around $\bm H$ with a uniform turn angle $\phi$ upon propagating along the $\bm H$ direction as shown in Fig.~\ref{Fig13}(c). In addition this state has a net magnetization parallel to $\bm H$. With the turn angle $\phi$ and the magnetization per site $m_z$=$m\cos\theta$, the magnetization vector on the $i$th plane and that on the ($i+1$)th plane are written, respectively, as
\begin{equation}
\bm m_i=m(\sin\theta \cos\phi_0, \sin\theta \sin\phi_0, \cos\theta),
\end{equation}
and
\begin{equation}
\bm m_{i+1}=m(\sin\theta \cos(\phi+\phi_0), \sin\theta \sin(\phi+\phi_0), \cos\theta).
\end{equation}
Then the energy per magnetization is given using $\theta$ and $\phi$ as,
\begin{eqnarray}
E[\theta, \phi]
&=&-Jm^2(\sin^2\theta \cos\phi + \cos^2\theta)
\nonumber \\
& &-Dm^2\sin^2\theta \sin\phi
\nonumber \\
& &-g\mu_{\rm B} \mu_0 H_zm\cos\theta.
\end{eqnarray}
The expressions of $\theta$ and $\phi$ can be derived from the saddle-point equations of the energy with respect to $\theta$ and $\phi$:
\begin{equation}
\frac{\partial E[\theta, \phi]}{\partial \theta}=0,
\;\;
\frac{\partial E[\theta, \phi]}{\partial \phi}=0.
\end{equation}
They are given by
\begin{eqnarray}
\phi&=&\tan^{-1} \left(\frac{D}{J} \right),
\label{eq:theta}
\\
\cos\theta&=&\frac{g\mu_{\rm B} \mu_0 H_z}{2m[J(\cos\phi-1)+D\sin\phi]}.
\label{eq:cosphi}
\end{eqnarray}
Note that if the magnitude of the external $\bm H$ is small enough, the propagation vector $\bm q$ for the conical state is not necessarily directed along $\bm H$ in reality, but could be pinned in a certain direction due to weak magnetic anisotropies. However it is assumed here that the vector $\bm q$ is always parallel to $\bm H$ for simplicity.

In the lower two panels of Fig.~\ref{Fig13}(b), calculated $H$-dependence of [110] component of the net magnetization $M_{[110]}$ and [001] component of the ferroelectric polarization $P_{[001]}$ under $\bm H$$\parallel$[110] are plotted for the three-dimensional case. Here the $M_{[110]}$ data indicated by closed circles are calculated for the ground-state magnetic configurations obtained by numerically minimizing the energy, while the analytically obtained behavior of $M_z$=$\cos \theta$ with Eq.~(\ref{eq:theta}) and Eq.~(\ref{eq:cosphi}) is indicated by a solid line. They show perfect coincidence.

\section{Electromagnetism in Multiferroics}
\label{sec4}

\subsection{Dynamical susceptibilities}
\begin{figure}
\begin{center}
\includegraphics[width=1.0\columnwidth]{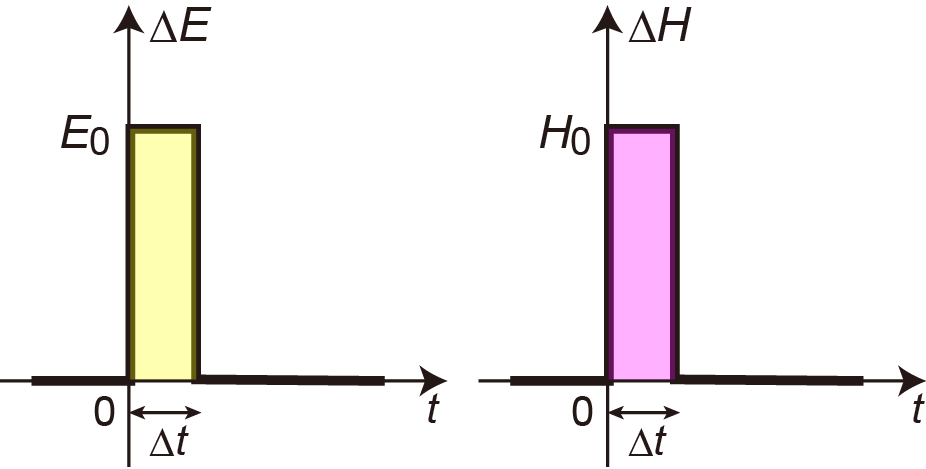}
\end{center}
\caption{(color online). Short rectangular pulses of magnetic field and electric field used in the numerical simulations for calculating the dynamical susceptibilities.}
\label{Fig14}
\end{figure}
Dynamical properties of resonantly oscillating magnetizations and magnetically induced polarizations of multiferroic skyrmions are captured by the following dynamical susceptibilities:\\
\\
{\bf Dynamical dielectric susceptibilities:}
\begin{eqnarray}
\chi^{\rm ee}_{\alpha \beta}(\omega) =
\frac{\Delta P_{\alpha}^{\omega}}{\epsilon_0 E_{\beta}^{\omega}}
\label{eq:EPsuscept}
\end{eqnarray}
{\bf Dynamical magnetic susceptibilities:}
\begin{eqnarray}
\chi^{\rm mm}_{\alpha \beta}(\omega) =
\frac{\Delta M_{\alpha}^{\omega}}{\mu_0 H_{\beta}^{\omega}}
\label{eq:HMsuscept}
\end{eqnarray}
{\bf Dynamical magnetoelectric susceptibilities:}
\begin{eqnarray}
\chi^{\rm em}_{\alpha \beta}(\omega) =
\frac{\Delta P_{\alpha}^{\omega}}{\sqrt{\epsilon_0 \mu_0}H_{\beta}^{\omega}}
=c\frac{\Delta P_{\alpha}^{\omega}}{H_{\beta}^{\omega}}
\label{eq:HPsuscept}
\end{eqnarray}
{\bf Dynamical electromagnetic susceptibilities:}
\begin{eqnarray}
\chi^{\rm me}_{\alpha \beta}(\omega) =
\sqrt{\frac{\mu_0}{\epsilon_0}}
\frac{\Delta M_{\alpha}^{\omega}}{E_{\beta}^{\omega}}
=c\mu_0 \frac{\Delta M_{\alpha}^{\omega}}{E_{\beta}^{\omega}}.
\label{eq:EMsuscept}
\end{eqnarray}
The latter two susceptibilities describe cross-correlation responses, that is, responses of polarization $\bm P$ to the ac magnetic field $\bm H^\omega$ and responses of magnetization $\bm M$ to the ac electric field $\bm E^\omega$. Here the indices $\alpha$ and $\beta$ run over the Cartesian coordinates.

These dynamical susceptibilities are calculated numerically using the Landau-Lifshitz-Gilbert equation:
\begin{equation}
\frac{d \bm m_i}{d t}=-\bm m_i \times \bm B^{\rm eff}_i
+ \frac{\alpha_{\rm G}}{m} \bm m_i \times \frac{d \bm m_i}{d t},
\label{eq:LLGEQ1}
\end{equation} 
where the local magnetization $\bm m_i$ is defined as $\bm m_i=-\bm S_i/\hbar$ with $\bm S_i$ being the spin. Here the first term depicts the gyrotropic motion of magnetization $\bm m_i$ under the effective magnetic field $\bm B^{\rm eff}_i$, while the second term describes the phenomenologically introduced damping effect with $\alpha_{\rm G}$ being the Gilbert-damping coefficient. After the linearization, the equation leads,
\begin{equation}
\frac{d \bm m_i}{d t}=\frac{1}{1+\alpha_{\rm G}^2} \left[
-\bm m_i \times \bm B^{\rm eff}_i - \frac{\alpha_{\rm G}}{m}
\bm m_i \times (\bm m_i \times \bm B^{\rm eff}_i) \right].
\label{eq:LLGEQ2}
\end{equation} 
The effective magnetic field $\bm B^{\rm eff}_i$ acting on the magnetization $\bm m_i$ is calculated from the derivative of the Hamiltonian with respect to $\bm m_i$ as,
\begin{equation}
\bm B^{\rm eff}_i = - \partial \mathcal{H} / \partial \bm m_i.
\label{eq:Heff}
\end{equation} 
with
\begin{eqnarray}
\mathcal{H}=\mathcal{H}_0 + \mathcal{H}^{\prime}(t).
\label{eq:Hamilt1}
\end{eqnarray} 
Here the first term $\mathcal{H}_0$ depicts the magnetic system of the chiral-lattice magnet, which is given by,
\begin{eqnarray}
\mathcal{H}_0&=& 
-J \sum_{<i,j>} \bm m_i \cdot (\bm m_{i+\hat{x}}+\bm m_{i+\hat{y}}) 
\nonumber \\
& &-D \sum_{i,\hat{\gamma}} \bm m_i \times \bm m_{i+\hat{\gamma}} \cdot \hat{\gamma}
\nonumber \\
& &-g\mu_{\rm B}\mu_0 \bm H \cdot \sum_i \bm m_i,
\label{eq:Hamilt2}
\end{eqnarray}
where $\bm H$=(0, 0, $H_z$)($\parallel$$\bm z$), and $\hat{\bm \gamma}$ runs over the Cartesian coordinates $\hat{\bm x}$, $\hat{\bm y}$ and $\hat{\bm z}$. The second term $\mathcal{H}^{\prime}(t)$ represents a short rectangular pulse of magnetic field or that of electric field as a perturbation whose time width is $\Delta t$:
\begin{eqnarray}
\mathcal{H}^{\prime}(t)=-g\mu_{\rm B} \mu_0 \sum_i \Delta \bm H(t) \cdot \bm m_i,
\label{eqn:rectpulseH}
\end{eqnarray}
or
\begin{eqnarray}
\mathcal{H}^{\prime}(t)=-\sum_i \Delta \bm E(t) \cdot \bm p_i.
\label{eqn:rectpulseE}
\end{eqnarray}
Time profiles of $\Delta \bm H(t)$ and $\Delta \bm E(t)$ are shown in Fig.~\ref{Fig14}. An advantage of utilizing short rectangular pulses is that the $\omega$-dependence in the Fourier components $\bm H^{\omega}$ and $\bm E^{\omega}$ shows up only in higher-order terms with respect to $\omega \Delta t$ as,
\begin{eqnarray}
E_{\alpha}^{\omega}=\int_{-\infty}^{\infty} \Delta E(t) e^{i\omega t} dt
=E_0 \Delta t + O[(\omega \Delta t)^2],
\label{eqn:Eomega}
\\
H_{\alpha}^{\omega}=\int_{-\infty}^{\infty} \Delta H(t) e^{i\omega t} dt
=H_0 \Delta t + O[(\omega \Delta t)^2].
\label{eqn:Homega}
\end{eqnarray}
Since the leading terms of $E_{\alpha}^{\omega}$ and $H_{\alpha}^{\omega}$ are $\omega$-independent, one only needs to calculate $\Delta M_{\alpha}^{\omega}$ and $\Delta P_{\alpha}^{\omega}$ to evaluate the dynamical susceptibilities given by Eqs.~(\ref{eq:EPsuscept})-(\ref{eq:EMsuscept}).
In the simulations, one should first trace the time evolution of the net magnetization,
\begin{eqnarray}
\bm M(t)=\frac{g\mu_{\rm B}}{NV} \sum_i \bm m_i(t),
\label{eqn:Mt}
\end{eqnarray}
and that of the ferroelectric polarization,
\begin{eqnarray}
\bm P(t)=\frac{1}{NV} \sum_i \bm p_i(t),
\label{eqn:Pt}
\end{eqnarray}
after application of a short pulse of $\mathcal{H}^{\prime}(t)$ at $t$=0. Then one performs the Fourier transformation of $\Delta \bm M(t)(=\bm M(t)-\bm M(t=0))$ and $\Delta \bm P(t)(=\bm P(t)-\bm P(t=0))$ to obtain $\Delta M_{\alpha}^{\omega}$ and $\Delta P_{\alpha}^{\omega}$.

\subsection{Spin-wave modes of skyrmion crystal: Theory}
\begin{figure}
\begin{center}
\includegraphics[width=1.0\columnwidth]{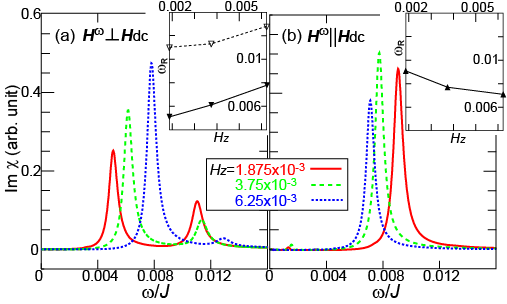}
\end{center}
\caption{(color online). (a) Frequency dependence of imaginary part of the dynamical magnetic susceptibilities, Im$\chi^{\rm mm}(\omega)$, in the skyrmion-crystal phase with several values of magnetic field $H_z$ for in-plane ac magnetic fields $\bm H^\omega$$\perp$$\bm H_{\rm dc}$. (b) That for out-of-plane ac magnetic fields $\bm H^\omega$$\parallel$$\bm H_{\rm dc}$. The static magnetic field $\bm H_{\rm dc}$=(0, 0, $H_z$) is applied normal to the two-dimensional plane. The angular frequency $\omega$ is normalized by the ferromagnetic-exchange coupling $J$, and $\omega/J$=0.01 corresponds to $\sim$1 GHz when $J$=0.4 meV for example. Insets show resonance frequencies $\omega_{\rm R}$ as functions of $H_z$. (Reproduced from Ref.~\cite{Mochizuki12}.)}
\label{Fig15}
\end{figure}
It was theoretically revealed that the skyrmion-crystal state has peculiar spin-wave modes with frequencies of a few gigahertz~\cite{Mochizuki12,Petrova11,Moutafis09,Makhfudz12,LinSZ14b,DaiY14,Tatara14}. Shown in Figs.~\ref{Fig15}(a) and (b) are calculated frequency dependence of imaginary part of the dynamical magnetic susceptibility Im$\chi^{\rm mm}(\omega)$ for the in-plane ac magnetic field $\bm H^\omega$$\perp$$\bm H_{\rm dc}$ and that for the out-of-plane ac magnetic field $\bm H^\omega$$\parallel$$\bm H_{\rm dc}$, respectively~\cite{Mochizuki12}. The calculations are performed by numerically solving the Landau-Lifshitz-Gilbert equation with the lattice spin model given by Eq.~(\ref{eq:Hamilt2}) in two dimensions where the static magnetic field $\bm H_{\rm dc}$ is applied normal to the plane. We find that each spectrum in Fig.~\ref{Fig15}(a) has a couple of resonance peaks, while that in Fig.~\ref{Fig15}(b) has only one resonance peak.

In order to identify these resonant modes, numerical simulations have been performed to trace the spatiotemporal dynamics of the local magnetizations $\bm m_i$ and the local scalar spin chiralities $C_i$ for the skyrmion-crystal state by applying a stationary oscillating ac magnetic field $\bm H^\omega$ whose frequency coincides with the resonance frequency of the corresponding mode~\cite{Mochizuki12}. Here the local scalar spin chirality is defined as,
\begin{eqnarray}
C_i=\bm m_i \cdot (\bm m_{i+\hat{x}} \times \bm m_{i+\hat{y}})
+\bm m_i \cdot (\bm m_{i-\hat{x}} \times \bm m_{i-\hat{y}}).
\label{eqn:pSDMLH}
\end{eqnarray}
It is found that for every resonance mode discussed here, all of the skyrmions exhibit uniformly the same motion so that one only needs to focus on the dynamics of one skyrmion in the skyrmion crystal to identify the microwave-active eigenmodes.

\begin{figure*}
\includegraphics[width=2.0\columnwidth]{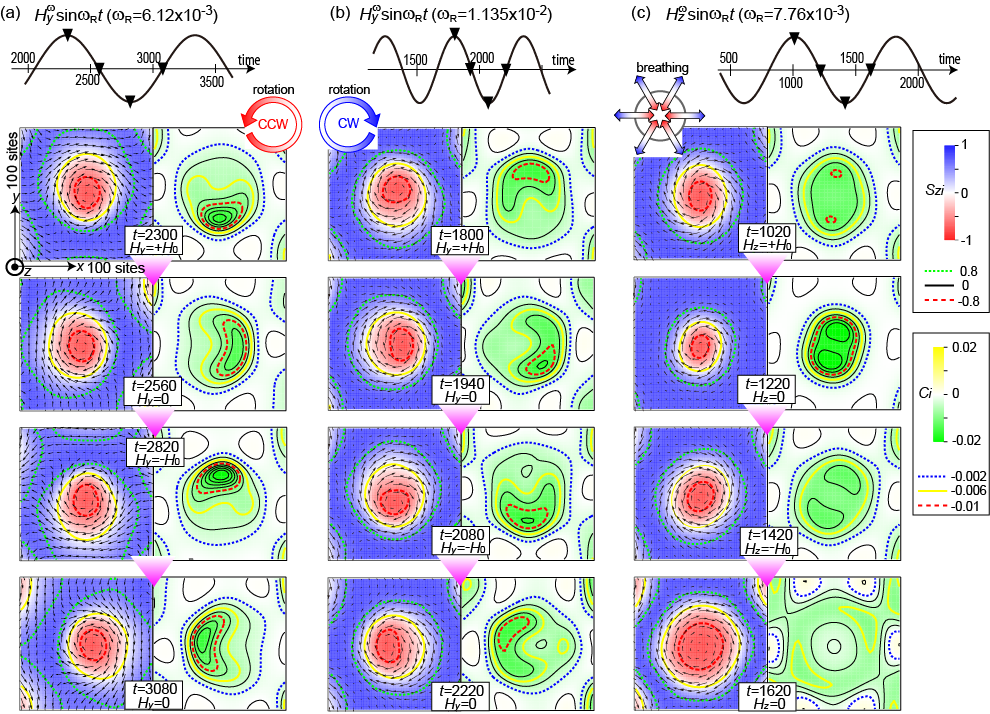}
\caption{(color online). Simulated spatiotemporal dynamics of local magnetizations (left panels) and local scalar spin chiralities $C_i$ (right panels) for three different spin-wave modes of the skyrmion crystal. Temporal waveforms of the applied ac magnetic fields are shown in the uppermost panels where inverted triangles indicate times at which we observe the snapshots shown here. (a) [(b)] Lower-energy [Higher-energy] rotational mode activated by the ac magnetic field $\bm H^\omega$ normal to the static magnetic field $\bm H_{\rm dc}$ ($\bm H^\omega$$\perp$ $\bm H_{\rm dc}$). Distributions of the $m_{zi}$ components and the spin chiralities $C_i$ circulate around the skyrmion core in a counterclockwise (clockwise) fashion. (c) Breathing mode activated the ac magnetic field $\bm H^\omega$ parallel to the static magnetic field $\bm H_{\rm dc}$ ($\bm H^\omega$$\parallel$ $\bm H_{\rm dc}$). Areas of the skyrmions expand and shrink in an oscillatory manner. (Reproduced from Ref.~\cite{Mochizuki12}.)}
\label{Fig16}
\end{figure*}
In Fig.~\ref{Fig16}, snapshots of the simulated spatiotemporal dynamics of a skyrmion in the collectively oscillating skyrmion crystal for each mode are displayed. As shown in Figs.~\ref{Fig16}(a) and (b), two resonant modes under $\bm H^\omega$$\perp$$\bm H_{\rm dc}$ are rotational modes where distribution of the out-of-plane magnetization components circulates around each skyrmion core. Interestingly such rotational motions of skyrmions are driven even by a linearly polarized $\bm H^\omega$. The rotation sense of the skyrmion circulation is opposite between these two modes. It is counterclockwise with respect to the $\bm H_{\rm dc}$ direction for the lower-frequency mode, while clockwise for the higher-frequency mode. The rotation sense becomes opposite when one reverses the sign of $\bm H_{\rm dc}$ or the sign of the skyrmion number $Q$. In contrast, it is not affected by the sign of the Dzyaloshinskii-Moriya parameter or swirling direction of the magnetizations. We find that intensity of the lower-lying mode is much stronger than the higher-lying mode because the rotation sense of the lower-lying mode matches the precessional direction of magnetizations determined by the sign of $\bm H_{\rm dc}$. On the other hand, the single resonant mode under $\bm H^\omega$$\parallel$$\bm H_{\rm dc}$ turns out to be a breathing mode where all the skyrmions in the skyrmion crystal expand and shrink in an oscillatory manner as shown in Fig.~\ref{Fig16}(c).

The spin-wave excitations activated by the in-plane ac magnetic field strongly depend on circular polarization of $\bm H^\omega$ or that of an irradiated microwave because of their rotational habits~\cite{Mochizuki12}. Numerical simulation found that irradiation of the left-handed circularly polarized microwave with resonant frequency significantly enhances the magnetization oscillation of the lower-lying counterclockwise rotational mode, whereas the right-handed circularly polarized microwave with the same frequency cannot activate the magnetization oscillation so much. Numerical simulations also demonstrated that melting of the skyrmion crystal can be realized when the counterclockwise rotational spin-wave mode is intensely excited in the skyrmion-crystal phase with the left-handed circularly polarized microwave. The melting occurs as the radius of the skyrmion-core rotation exceeds the lattice constant of the skyrmion crystal.

\subsection{Spin-wave modes of skyrmion crystal: Experiment}
\begin{figure*}
\begin{center}
\includegraphics[width=2.0\columnwidth]{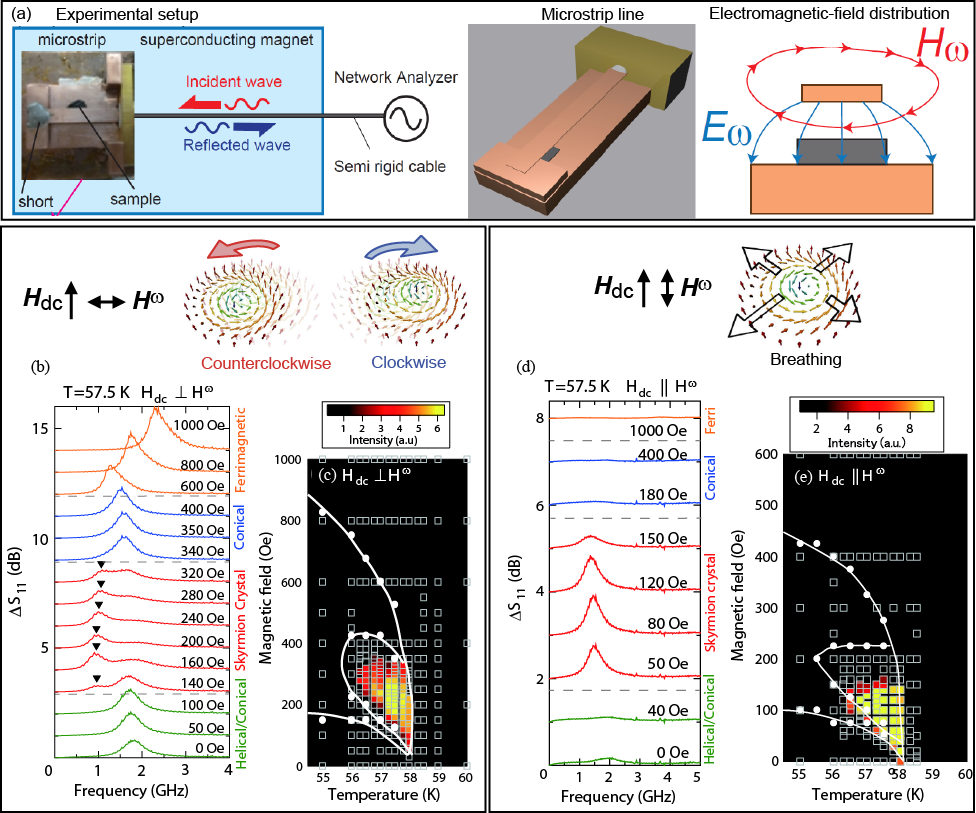}
\end{center}
\caption{(color online). (a) Setup of the microwave absorption experiment (left panel), the microstrip line used to detect the magnetic resonant modes of the skyrmion crystal (middle panel), and the electromagnetic-field distribution in the present setup using the microstrip line viewed along the microwave propagation direction (right panel). (b) Experimentally measured microwave absorption spectra under various magnitudes of static magnetic field $H_{\rm dc}$ at 57.5 K for bulk samples of Cu$_2$OSeO$_3$ and (c) temperature versus magnetic field phase diagram with background color indicating the absorption intensity of the skyrmion resonant modes for microwave-polarization configuration of $\bm H^\omega$$\perp$$\bm H_{\rm dc}$ with which clockwise and counterclockwise rotational modes are activated in the skyrmion-crystal phase. (d),(e) Corresponding profiles for microwave-polarization configuration of $\bm H^\omega$$\parallel$$\bm H_{\rm dc}$ with which a breathing mode is activated in the skyrmion-crystal phase. (Reproduced from Ref.~\cite{Onose12}.)}
\label{Fig17}
\end{figure*}
Since the magnon excitations are characterized by typical resonance frequencies ranging from gigahertz to terahertz, the measurement of absorption and/or transmission spectra with linearly polarized electromagnetic waves in this energy region is the most straightforward way to experimentally identify them. However the dominant contribution from the Drude response of conduction electrons often prevents the detection of pure magnetization dynamics in metallic systems. Therefore the employment of insulating materials is ideal for this purpose because of the absence of the Drude contribution. Indeed the predicted skyrmion resonant modes were detected by microwave absorption and transmission measurements for Cu$_2$OSeO$_3$~\cite{Onose12,Schwarze15}. The setup of the experiment in Ref.~\cite{Onose12} is shown in Fig.~\ref{Fig17}(a).

Figure~\ref{Fig17}(b) indicates measured absorption spectra for bulk samples of Cu$_2$OSeO$_3$ at 57.5 K (just below $T_{\rm c} \sim 58$ K) for various magnitudes of $H_{\rm dc}$ with a microwave polarization of $\bm H^\omega$$\perp$ $\bm H_{\rm dc}$. The helical spin order is realized for $H_{\rm dc} \leq 400$ Oe except for the region 140 Oe$ \leq H_{\rm dc} \leq 320$ Oe where the skyrmion-crystal phase takes place. While the helical spin state shows a magnetic resonance at around 1.5-1.7 GHz, the emergence of new absorption mode around 1 GHz is clearly observed in the latter $H_{\rm dc}$-range. In Fig.~\ref{Fig17}(c), intensity of the new absorption mode is indicated by colors in plane of $T$ and $H_{\rm dc}$. This mode has been confirmed to appear only in the skyrmion-crystal phase, and can be assigned to the counterclockwise rotational mode of the skyrmion crystal according to the theoretically suggested selection rules with respect to the microwave polarization. Note that the existence of the clockwise rotational mode at higher resonance frequency was also predicted for the skyrmion-crystal state, while its experimental identification is yet to be performed because of its weak absorption intensity.

Likewise the corresponding $H_{\rm dc}$-dependence of the absorption spectra for the microwave polarization $\bm H^\omega$$\parallel$$\bm H_{\rm dc}$ is shown in Fig.~\ref{Fig17}(d). Whereas the conventional ferromagnetic resonance is active only to $\bm H^\omega$($\perp$ $\bm H_{\rm dc}$) and this selection rule holds also for the helical state, emergence of a sharp absorption peak is observed at 1.3-1.5 GHz for the region 50 Oe $\leq H_{\rm dc} \leq 150$ Oe. The color plot of the spectral intensity in the $H$-$T$ plane indicates that this mode uniquely appears in the skyrmion-crystal phase [Fig.~\ref{Fig17}(e)]. Comparison with the theoretically proposed selection rules suggests that this resonant absorption corresponds to the breathing mode of the skyrmion crystal.

\section{Microwave magnetoelectric phenomena}
\label{sec5}

\subsection{Electromagnetic waves in multiferroics}
Dynamical responses of multiferroic materials with net magnetization $\bm M$ and ferroelectric polarization $\bm P$ to an irradiated electromagnetic wave can be discussed starting from Maxwell's equations,
\begin{eqnarray}
{\rm rot}{\bm E} &=&-\frac{\partial \bm B}{\partial t},
\label{eqn:MaxwellEq1}
\\
{\rm rot}{\bm B} &=& \frac{\partial \bm D}{\partial t},
\label{eqn:MaxwellEq2}
\end{eqnarray}
where
\begin{eqnarray}
{\bm B} &=& \mu_0 (\hat{\mu}^\infty {\bm H}+{\bm M}),
\label{eqn:B}
\\
{\bm D} &=& \epsilon_0 \hat{\epsilon}^\infty {\bm E}+{\bm P},
\label{eqn:D}
\\
{\bm P} &=& \epsilon_0 \hat{\chi}^{\rm ee} {\bm E}
+ \hat{\chi}^{\rm em}  \sqrt{\epsilon_0 \mu_0} {\bm H},
\label{eqn:P}
\\
{\bm M} &=& \mu_0 \hat{\chi}^{\rm mm} {\bm H}
+ \hat{\chi}^{\rm me} \sqrt{\frac{\epsilon_0}{\mu_0}} {\bm E}.
\label{eqn:M}
\end{eqnarray}
Note that because of the multiferroic nature, the magnetoelectric susceptibilities $\hat{\chi}^{\rm em}$ and $\hat{\chi}^{\rm me}$ become finite, and there appear a contribution proportional to $\bm H$ in the expression of the ferroelectric polarization $\bm P$ and that proportional to $\bm E$ in the expression of the net magnetization $\bm M$. Inserting the following expressions,
\begin{eqnarray}
{\bm E}&=&{\bm E}^\omega \exp[{\rm i}({\bm K^\omega} \cdot {\bm r} - \omega t)],
\label{eqn:E}
\\
{\bm H}&=&{\bm H}^\omega \exp[{\rm i}({\bm K^\omega} \cdot {\bm r} - \omega t)],
\label{eqn:H}
\end{eqnarray}
into Eq.~(\ref{eqn:MaxwellEq1}) and Eq.~(\ref{eqn:MaxwellEq2}), one obtains the following equations for Fourier components,
\begin{eqnarray}
 \omega {\bm B}^\omega &=& {\bm K^\omega} \times {\bm E}^\omega,
\label{eqn:FTMaxwellEq1}
\\
-\omega {\bm D}^\omega &=& {\bm K^\omega} \times {\bm H}^\omega,
\label{eqn:FTMaxwellEq2}
\end{eqnarray}
where
\begin{eqnarray}
{\bm B}^\omega \!\! &=& \!\! \left[ \hat{\mu}^\infty 
+ \hat{\chi}^{\rm mm}(\omega) \right] \mu_0 {\bm H}^\omega 
+ \hat{\chi}^{\rm me}(\omega) \sqrt{\epsilon_0 \mu_0}  {\bm E}^\omega,
\label{eqn:InducedB}
\\
{\bm D}^\omega \!\! &=& \!\! \left[ \hat{\epsilon}^\infty 
+ \hat{\chi}^{\rm ee}(\omega) \right] \epsilon_0 {\bm E}^\omega 
+ \hat{\chi}^{\rm em}(\omega) \sqrt{\epsilon_0 \mu_0} {\bm H}^\omega.
\label{eqn:InducedD}
\end{eqnarray}
Solving these simultaneous equations in terms of the wave vector $\bm K^{\omega}$ and considering the relation,
\begin{eqnarray}
K^\omega=\omega N(\omega)/c,
\label{eqn:Komega}
\end{eqnarray}
one obtains an expression of the complex refractive index
\begin{eqnarray}
N(\omega)=n(\omega)+i\kappa(\omega).
\label{eqn:Nomega}
\end{eqnarray}
This quantity gives us information about how the material responds to an irradiated electromagnetic wave. In particular, its imaginary part $\kappa(\omega)$ is called extinction coefficient and describes to what extent the material absorbs the electromagnetic wave. More concretely, the absorption coefficient $\alpha(\omega)$ is related to $\kappa(\omega)$ as,
\begin{eqnarray}
\alpha(\omega)=\frac{2\omega\kappa(\omega)}{c} =\frac{2\omega}{c} {\rm Im}N(\omega),
\end{eqnarray}
and the absorption intensity $I(\omega)$ is given by,
\begin{eqnarray}
I(\omega)=I_0 \exp[-\alpha(\omega) d],
\end{eqnarray}
with $d$ being the sample thickness.

\subsection{Nonreciprocal directional dichroism: Theory}
In the multiferroic materials with magnetically induced electric polarizations, one can expect dynamical coupling of magnetic and electric dipoles. For example, resonant oscillations of magnetizations in multiferroics are coupled to those of electric polarizations owing to the magnetoelectric coupling, and thus can be activated not only magnetically by an ac magnetic field $\bm H^\omega$ but also electrically by an ac electric field $\bm E^\omega$~\cite{Smolenski82,Pimenov06,Katsura07}. Such simultaneous magnetic and electric activities of magnons (so-called electromagnons) cause intriguing dynamical magnetoelectric phenomena in the microwave frequency regime~\cite{Mochizuki13,Okamura13}. Since the magnetic skyrmions in Cu$_2$OSeO$_3$ are accompanied by electric dipoles, the resonant oscillation of skyrmion crystal active to $\bm H^\omega$ can be also activated by $\bm E^\omega$. The theoretical calculation indeed predicted that the rotational modes and the breathing mode of skyrmion crystal under $\bm H_{\rm dc}$$\parallel$[110] has finite electric susceptibility for $\bm E^\omega$$\parallel$[110]($\parallel$$\bm H_{\rm dc}$) and $\bm E^\omega$$\parallel$[001]($\perp$$\bm H_{\rm dc}$), respectively. 

\begin{figure}
\begin{center}
\includegraphics[width=1.0\columnwidth]{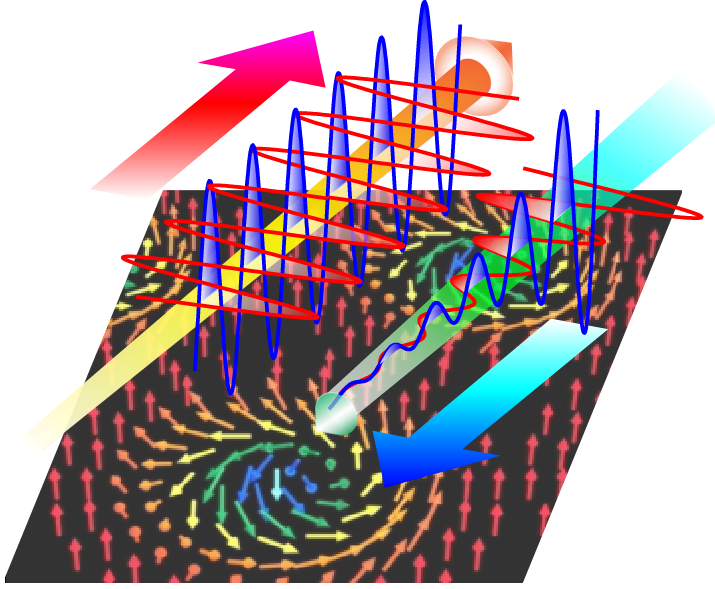}
\end{center}
\caption{(color online). Schematic illustration of the nonreciprocal directional dichroism of electromagnetic waves induced by the skyrmion-crystal state. An electromagnetic wave irradiating in a certain direction to a material is strongly absorbed, while that in the opposite direction is not absorbed so much.}
\label{Fig18}
\end{figure}
For such hybridized electromagnon modes, interference between the $\bm H^\omega$-activity and the $\bm E^\omega$-activity often leads to unique optical and/or microwave phenomena called directional dichroism~\cite{Mochizuki13}. This is a kind of "one-way window" effect where reversal of incident direction of an electromagnetic wave gives different absorption or emission spectrum (see schematic figure in Fig.~\ref{Fig18}). Since the first experimental demonstration by Rikken in 1997~\cite{Rikken97}, directional dichroism in absorption and emission has been reported for various frequency range from x-ray to visible-light to terahertz~\cite{Rikken02,Arima08,Kubota04,Jung04,Saito08a,Saito08b,Takahashi12,Bordacs12,Miyahara11}. The directional dichroism can be considered as a direct expansion of the linear magnetoelectric effect ($P_i = \alpha_{ij} H_j$ and $M_i = \alpha_{ji} E_j$) into the dynamical regime and is allowed when $\alpha_{xy}$ component is nonzero for the incident direction of an electromagnetic wave $\bm K^\omega \parallel \bm z$. This condition is always satisfied when the relationship $(\bm P \times \bm M) \parallel \bm K^\omega$ holds, while the magnitude of directional dichroism essentially depends on nature of the absorption at the target frequency.

\begin{figure}
\begin{center}
\includegraphics[width=1.0\columnwidth]{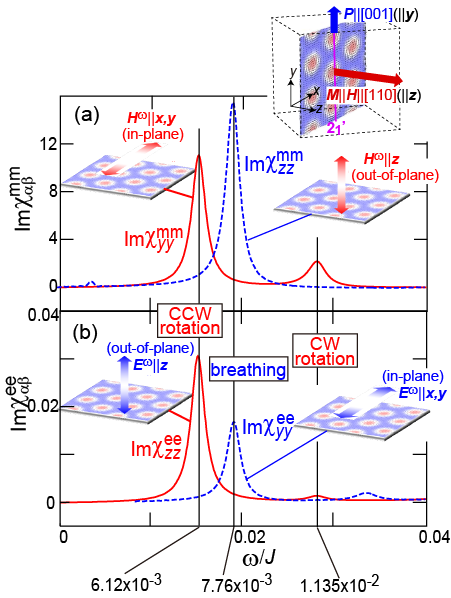}
\end{center}
\caption{(color online). Calculated spectra of (a) imaginary part of the dynamical magnetic susceptibility Im$\chi^{\rm mm}_{\alpha \beta}(\omega)$Cand (b) imaginary part of the dynamical dielectric susceptibility $\chi^{\rm ee}_{\alpha \beta}(\omega)$ for the skyrmion-crystal phase under the external magnetic field $\bm H$$\parallel$[110]. Im$\chi^{\rm mm}_{yy}$ in (a) and Im$\chi^{\rm ee}_{zz}$ in (b) have resonance peaks at the same frequencies, indicating that both the in-plane ac magnetic field $\bm H^\omega$$\parallel$$\bm y$ and the out-of-plane ac electric field $\bm E^\omega$$\parallel$$\bm z$ activate common eigenmodes. On the other hand, Im$\chi^{\rm mm}_{zz}$ in (a) and Im$\chi^{\rm ee}_{yy}$ in (b) also have a resonance peak at the same frequency, indicating that both the out-of-plane ac magnetic field $\bm H^\omega$$\parallel$$\bm z$ and the in-plane ac electric field $\bm E^\omega$$\parallel$$\bm y$ activate a common eigenmode. For the definition of the Cartesian coordinates $x$, $y$ and $z$, see the inset. (Reproduced from Ref.~\cite{Mochizuki13}.)}
\label{Fig19}
\end{figure}
When one applies a static magnetic field $\bm H_{\rm dc}$$\parallel$[110] to Cu$_2$OSeO$_3$, the ferroelectric polarization $\bm P$$\parallel$[001] as well as the net magnetization $\bm M$$\parallel$[110] are induced, which are orthogonal to each other as shown in Fig.~\ref{Fig07}(c) or Fig.~\ref{Fig09}(b). With this special configuration of $\bm P$$\perp$$\bm M$ , both $\bm H^\omega$ and $\bm E^\omega$ components of a microwave propagating parallel or antiparallel to $\bm P$$\times$$\bm M$ can excite common oscillation modes. 

Such a simultaneous activity of the collective modes can be seen via comparison between the dynamical magnetic susceptibilities $\chi^{\rm mm}_{\alpha \beta}(\omega)$ and the dynamical dielectric susceptibilities $\chi^{\rm ee}_{\alpha \beta}(\omega)$. Figures~\ref{Fig19}(a) and (b) show calculated frequency dependence of Im$\chi^{\rm mm}_{\alpha \beta}$ and that of Im$\chi^{\rm ee}_{\alpha \beta}$, respectively, for the skyrmion-crystal state under $\bm H$$\parallel$[110]. We find that resonances active to ac magnetic fields seen as peaks in the spectrum of Im$\chi^{\rm mm}_{\alpha \beta}$ can be found also in the spectrum of Im$\chi^{\rm ee}_{\alpha \beta}$. More concretely, the spectrum of Im$\chi^{\rm mm}_{yy}$ with $\bm y \parallel [001]$ in Fig.~\ref{Fig19}(a) has a couple of resonance peaks, and the spectrum of Im$\chi^{\rm ee}_{zz}$ with $\bm z \parallel [110]$ in Fig.~\ref{Fig19}(b) also has peaks at the same frequencies. This indicates that the in-plane ac magnetic field $\bm H^\omega$$\perp$$\bm H_{\rm dc}$ and the out-of-plane ac electric field $\bm E^\omega$$\parallel$$\bm H_{\rm dc}$ activate the same eigenmodes. On the other hand, both Im$\chi^{\rm mm}_{zz}$ in Fig.~\ref{Fig19}(a) and Im$\chi^{\rm ee}_{yy}$ in Fig.~\ref{Fig19}(b) have a single peak at the same frequency, indicating that the out-of-plane ac magnetic field $\bm H^\omega$$\parallel$$\bm H_{\rm dc}$ and the in-plane ac electric field $\bm E^\omega$$\perp$$\bm H_{\rm dc}$ activate the same eigenmode.

\begin{figure}
\begin{center}
\includegraphics[width=1.0\columnwidth]{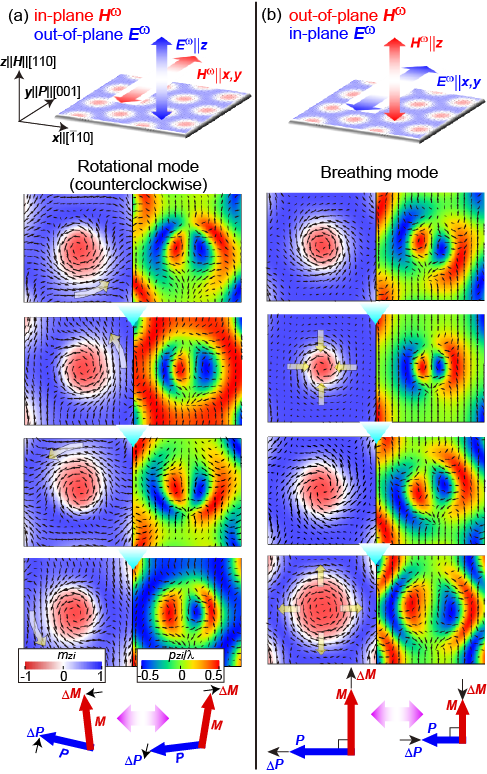}
\end{center}
\caption{(color online). Spatiotemporal dynamics of magnetizations $\bm m_i$ (left panels) and polarizations $\bm p_i$ (right panels) for electromagnon excitations in the skyrmion-crystal under the static magnetic field $\bm H$$\parallel$[110]($\parallel$$\bm z$). (a) Counterclockwise rotational mode activated by in-plane ac magnetic field $\bm H^\omega$$\perp$[110] or by out-of-plane ac electric field $\bm E^\omega$$\parallel$[110]. (b) Breathing mode activated by out-of-plane ac magnetic field $\bm H^\omega$$\parallel$[110] or by in-plane ac electric field $\bm E^\omega$$\perp$[110]. Dynamical behaviors of net magnetization $\bm M$ and ferroelectric polarization $\bm P$ are shown for each mode in the lowest panel. The angle between $\bm M$ and $\bm P$ become larger and smaller in an oscillatory manner in the rotational mode, while in the breathing mode, elongation and shrinkage of $\bm M$ and $\bm P$ in length occur synchronously, keeping the angle at 90$^\circ$.}
\label{Fig20}
\end{figure}
In Figs.~\ref{Fig20}(a) and (b), snapshots of the simulated coupled dynamics of local magnetizations $\bm m_i$ (left panels) and local polarizations $\bm p_i$ (right panels) are shown for the counterclockwise rotational mode activated by $\bm H^{\omega}$$\perp$$\bm H_{\rm dc}$ or $\bm E^{\omega}$$\parallel$$\bm H_{\rm dc}$ and the breathing mode activated by $\bm H^{\omega}$$\parallel$$\bm H_{\rm dc}$ or $\bm E^{\omega}$$\perp$$\bm H_{\rm dc}$, respectively. Behaviors of the net magnetization $\bm M$ and the ferroelectric polarization $\bm P$ for these modes are shown in the bottom panels. In the rotational mode, the angle between $\bm P$ and $\bm M$, which is originally 90$^{\circ}$, becomes larger and smaller oscillatorily. On the other hand, in the breathing mode, simultaneous elongation and shrinkage of $\bm P$ and $\bm M$ occur in an oscillatory manner, while keeping the angle between them 90$^{\circ}$. 

\begin{figure}
\begin{center}
\includegraphics[width=1.0\columnwidth]{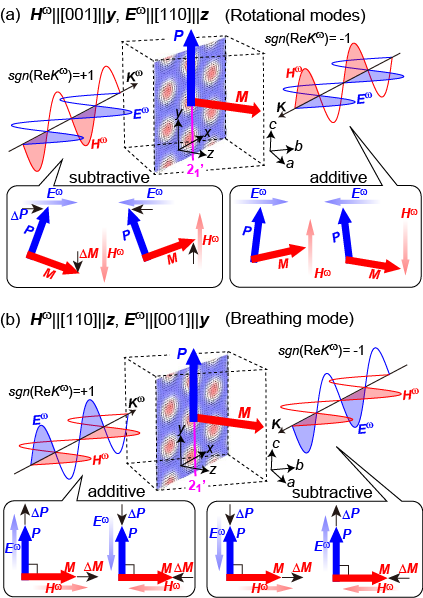}
\end{center}
\caption{(color online). Microwave polarization configurations with which the nonreciprocal directional dichroism occurs in the skyrmion-crystal phase with net magnetization $\bm M$$\parallel$[110]($\parallel$$\bm z$) and ferroelectric polarization $\bm P$$\parallel$[001]($\parallel$$\bm y$). (a) $\bm H^\omega$$\parallel$[001]($\parallel$$\bm y$) and $\bm E^\omega$$\parallel$[001]($\parallel$$\bm z$). The $\bm H^\omega$ and $\bm E^\omega$ components contribute in an additive way to the excitation of rotational modes for a microwave propagating in the negative ($-\bm x$ or [1$\bar{1}$0]) direction, while in a subtractive way for a microwave propagating in the positive ($+\bm x$ or [$\bar{1}$10]) direction. Consequently the absorption intensity of the former microwave with ${\rm sgn}(ReK^\omega)=-1$ becomes larger than that of the latter one with ${\rm sgn}(ReK^\omega)=+1$. (b) $\bm H^\omega$$\parallel$[110]($\parallel$$\bm z$) and $\bm E^\omega$$\parallel$[001]($\parallel$$\bm z$). In this case, a microwave propagating in the $+\bm x$ direction is absorbed intensely through strongly exciting the breathing mode as compared to that propagating in the $-\bm x$ direction. (Reproduced from Ref.~\cite{Mochizuki13}.)}
\label{Fig21}
\end{figure}
These simultaneous electric and magnetic activities of the resonant modes cause peculiar dynamical magnetoelectric phenomena of skyrmions in the microwave-frequency regime~\cite{Mochizuki13}. We first discuss the case with a linearly polarized microwave with $\bm H^{\omega}$$\parallel$[001]($\perp$$\bm M$) and $\bm E^{\omega}$$\parallel$[110]($\perp$$\bm P$) as shown in Fig.~\ref{Fig21}(a). Starting from Maxwell's equations, one can derive a relation for electromagnetic waves:
\begin{eqnarray}
\bm H^{\omega} \parallel \bm K^\omega \times \bm E^{\omega}.
\label{eqn:pSDMLH}
\end{eqnarray}
This equation indicates that the relative relationship between $\bm H^{\omega}$- and $\bm E^{\omega}$-directions of an electromagnetic wave is determined by the sign of the wave vector $\bm K^\omega$ or the propagation direction. This relative relationship will be reversed upon the sign reversal of $\bm K^\omega$. Accordingly the $\bm H^{\omega}$ and $\bm E^{\omega}$ components of a microwave propagating in [$\bar{1}$10] ($+\bm x$) direction cooperatively induce an oscillation of the angle between $\bm P$ and $\bm M$ so as to intensely activate the rotation mode [see bottom panels of Fig.~\ref{Fig21}(a)]. In turn, when the microwave is propagating in the opposite direction, that is, [1$\bar{1}$0] ($-\bm x$) direction, its $\bm H^{\omega}$ and $\bm E^{\omega}$ components work destructively to activate the rotation mode. Consequently the microwave with $\bm K^\omega$$\parallel$$-\bm x$ strongly excites the spin wave resonance, and thereby is strongly absorbed, while the microwave with $\bm K^\omega$$\parallel$$+\bm x$ excites the spin wave resonance only weakly, and thus is weakly absorbed. In this way, the microwave absorption intensity becomes different depending on the incident direction.

One can also expect that interference between the magnetic and electric activation processes of the breathing mode also gives rise to the nonreciprocal directional dichroism of microwave when the microwave polarization is $\bm H^{\omega}$$\parallel$[110]($\parallel$$\bm M$) and $\bm E^{\omega}$$\parallel$[001]($\parallel$$\bm P$) as shown in Fig.~\ref{Fig21}(b). In this case, the $\bm H^{\omega}$ and $\bm E^{\omega}$ components of a microwave propagating in the $+\bm x$ direction cooperatively induce length oscillations of $\bm P$ and $\bm M$ so as to intensely activate the breathing mode, whereas those of a microwave propagating in the $-\bm x$ direction do not [see bottom panels of Fig.~\ref{Fig21}(b)]. Consequently, the microwave with $\bm K^\omega$$\parallel$$+\bm x$ is strongly absorbed, whereas the microwave with $\bm K^\omega$$\parallel$$-\bm x$ is weakly absorbed. Again the absorption intensity becomes different depending on the incident direction of the microwave.

\begin{figure*}
\begin{center}
\includegraphics[width=2.0\columnwidth]{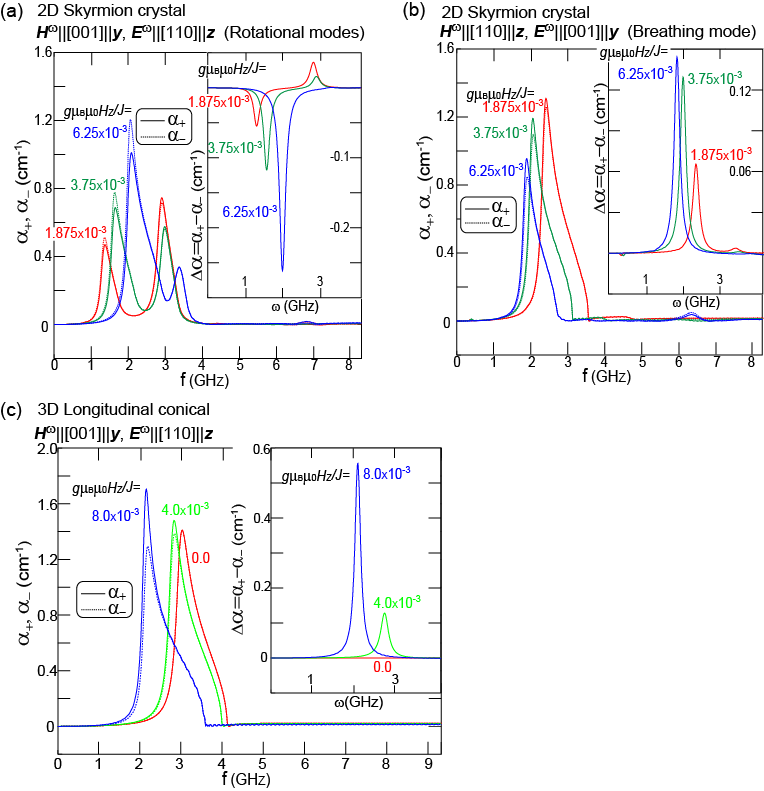}
\end{center}
\caption{(color online). (a),(b) Calculated microwave absorption spectra for the skyrmion-crystal phase realized in a (two-dimensional) thin-film specimen of Cu$_2$OSeO$_3$ under the external magnetic field $\bm H_{\rm dc}$$\parallel$[110]. Here $\alpha_+$ and $\alpha_-$ are the microwave absorption coefficients for positive and negative incident directions, that is, ${\rm sgn}({\rm Re}K^\omega)=+1$ and ${\rm sgn}({\rm Re}K^\omega)=-1$, respectively. The difference $\Delta \alpha$ represents the magnitude of the directional dichroism. The spectra in (a) are obtained for the microwave polarization of $\bm H^\omega \parallel [001] (\parallel \bm y)$ and $\bm E^\omega \parallel [110] (\parallel \bm z)$ where the rotational modes are excited. On the other hand, the spectra in (b) are obtained for the microwave polarization of $\bm H^\omega \parallel [110] (\parallel \bm z)$ and $\bm E^\omega \parallel [001] (\parallel \bm y)$ where the breathing mode is excited. (c) Calculated microwave absorption spectra $\alpha_+$ and $\alpha_-$ for the longitudinal conical phase in a (three-dimensional) bulk specimen of Cu$_2$OSeO$_3$ under $\bm H_{\rm dc}$$\parallel$[110] for $\bm H^\omega$$\parallel$[001]($\parallel$$\bm y$) and $\bm E^\omega$$\parallel$[110]($\parallel$$\bm z$). (Reproduced from Ref.~\cite{Mochizuki13}.)}
\label{Fig22}
\end{figure*}
For quantitative discussion on these effects, one derives expressions of a complex refractive index $N(\omega)$ by solving the simultaneous equations (\ref{eqn:FTMaxwellEq1}) and (\ref{eqn:FTMaxwellEq2}) for two different cases of microwave polarizations:\\
\\
Case (i): $\bm H^{\omega}$$\parallel$[001]($\parallel$$\bm y$) and $\bm E^{\omega}$$\parallel$[110]($\parallel$$\bm z$)
\begin{eqnarray}
N(\omega) &\sim& \sqrt{
[\epsilon_{zz}^\infty + \chi^{\rm ee}_{zz} (\omega)]
[\mu_{yy}^\infty + \chi^{\rm mm}_{yy} (\omega)]}
\nonumber \\
& &-{\rm sgn}({\rm Re}K^\omega) [\chi^{\rm me}_{yz} (\omega)
+ \chi^{\rm em}_{zy} (\omega)]/2,
\label{eqn:CRIHyEz}
\end{eqnarray}
\\
Case (ii): $\bm H^{\omega}$$\parallel$[110]($\parallel$$\bm z$) and $\bm E^{\omega}$$\parallel$[001]($\parallel$$\bm y$)
\begin{eqnarray}
N(\omega) &\sim& \sqrt{
[\epsilon_{yy}^\infty + \chi^{\rm ee}_{yy} (\omega)]
[\mu_{zz}^\infty + \chi^{\rm mm}_{zz} (\omega)]}
\nonumber \\
& &+ {\rm sgn}({\rm Re}K^\omega) [\chi^{\rm me}_{zy} (\omega)
+ \chi^{\rm em}_{yz} (\omega)]/2.
\label{eqn:CRIHzEy}
\end{eqnarray}
In these expressions, we find that there appears a term containing the sign of $K^{\omega}$. Since the absorption coefficient of electromagnetic waves is given by
$\alpha(\omega)=(2\omega/c) {\rm Im}N(\omega)$, the coefficient $\alpha(\omega)$ depends on the sign of $K^{\omega}$ or the incident direction of electromagnetic wave. Magnitude of the nonreciprocal directional dichroism or change of the absorption coefficient accompanied by reversal of the microwave incident direction, that is, $\Delta\alpha(\omega)=\alpha_{+}(\omega) - \alpha_{-}(\omega)$, is given by:\\
\\
Case (i): $\bm H^{\omega}$$\parallel$[001]($\parallel$$\bm y$) and $\bm E^{\omega}$$\parallel$[110]($\parallel$$\bm z$)
\begin{eqnarray}
\Delta\alpha(\omega)=
-{\rm Im}[\chi^{\rm me}_{yz} (\omega) + \chi^{\rm em}_{zy} (\omega)]/2,
\end{eqnarray}
\\
Case (ii): $\bm H^{\omega}$$\parallel$[110]($\parallel$$\bm z$) and $\bm E^{\omega}$$\parallel$[001]($\parallel$$\bm y$)
\begin{eqnarray}
\Delta\alpha(\omega)=
{\rm Im}[\chi^{\rm me}_{zy} (\omega) + \chi^{\rm em}_{yz} (\omega)]/2.
\end{eqnarray}
Here $\alpha_+(\omega)$ and $\alpha_-(\omega)$ are absorption coefficients for an electromagnetic wave propagating in the positive and negative $x$ directions, respectively. We find that the magnitude of nonreciprocal directional dichroism is governed by magnetoelectric susceptibilities
\begin{math}
\chi^{\rm em}_{\alpha \beta}(\omega)
\end{math}
and
\begin{math}
\chi^{\rm me}_{\alpha \beta}(\omega).
\end{math}

\begin{figure*}
\begin{center}
\includegraphics[width=2.0\columnwidth]{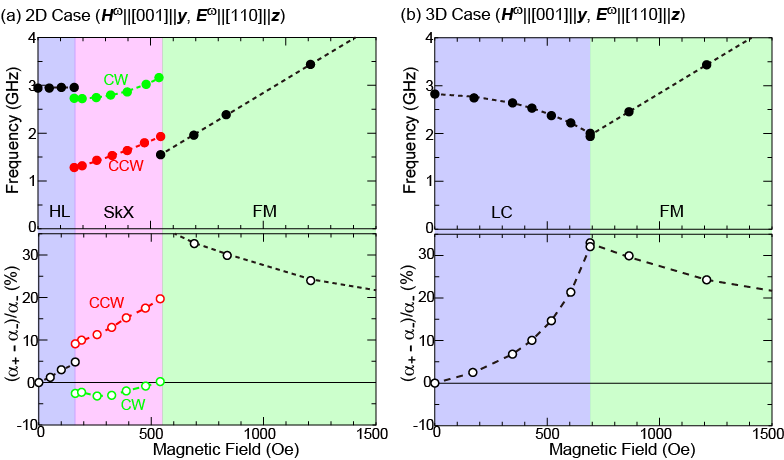}
\end{center}
\caption{(color online). Calculated $H_{\rm dc}$ dependence of the resonant frequency $f_{\rm R}$ (upper panels) and the magnitude of nonreciprocal directional dichroism $\Delta \alpha/\alpha_-$ (lower panels) for (a) two-dimensional system (thin-film case) and (b) three-dimensional system (bulk case) under $\bm H_{\rm dc} \parallel [110]$ for the microwave polarization $\bm H^{\omega} \parallel [001] (\perp \bm H_{\rm dc})$ and $\bm E^{\omega} \parallel [110] (\parallel \bm H_{\rm dc})$.}
\label{Fig23}
\end{figure*}
Figures~\ref{Fig22}(a) and (b) display calculated spectra of microwave absorption coefficients $\alpha_+(\omega)$ and $\alpha_-(\omega)$ as well as their difference $\Delta \alpha(\omega)$ for the skyrmion-crystal state in a (two-dimensional) thin-film sample of Cu$_2$OSeO$_3$ under $\bm H_{\rm dc}$$\parallel$[110] for several values of $H_{\rm dc}$. In Fig.~\ref{Fig22}(a)Cthere appear two peaks in the spectra, which correspond to the low-energy counterclockwise and the high-energy clockwise rotational modes, respectively, for the microwave polarization of $\bm H^\omega$$\parallel$[001] ($\parallel$$\bm y$) and $\bm E^\omega$$\parallel$[110] ($\parallel$$\bm z$). The difference of absorption coefficients $\Delta \alpha(\omega)$ becomes larger as the magnetic field $H_{\rm dc}$ increases, and eventually reaches a maximum value of 0.25 cm$^{-1}$, which corresponds to the relative difference of $\Delta\alpha$/$\alpha_{\rm ave}$=2($\alpha_+ - \alpha_-$)/($\alpha_+ + \alpha_-$)$\sim$20 $\%$. On the other hand, in Fig.~\ref{Fig22}(b), a single resonance peak, which corresponds to the breathing mode appears for the microwave polarization configuration of $\bm H^\omega \parallel [110] (\parallel \bm z)$ and $\bm E^\omega \parallel [001] (\parallel \bm y)$. The difference of absorption coefficients $\Delta \alpha(\omega)$ increases as the magnetic field $H_{\rm dc}$ increases, and reaches 0.14 cm$^{-1}$ at maximum which corresponds to the relative difference of $\sim$10 $\%$.

The microwave directional dichroism can be expected also in other magnetic phases. Figure~\ref{Fig22}(c) display calculated spectra of $\alpha_+(\omega)$ and $\alpha_-(\omega)$ as well as their difference $\Delta \alpha(\omega)$ for the longitudinal conical state in a (three-dimensional) bulk sample of Cu$_2$OSeO$_3$ under $\bm H_{\rm dc}$$\parallel$[110] for several values of $H_{\rm dc}$. The magnitude of $\Delta \alpha(\omega)$ increases as $H_{\rm dc}$ increases, and reaches relative difference of $\sim$30 $\%$. In Fig.~\ref{Fig23}(a) and (b), calculated resonant frequency $f_{\rm R}$ (upper panels) and magnitude of directional dichroism $\Delta \alpha/\alpha_-$ (lower panels) are plotted as functions of dc magnetic field $H_{\rm dc}$ for the two-dimensional case (thin-film sample) and the three-dimensional case (bulk sample) for the microwave polarization $\bm H^{\omega}$$\parallel$[001] ($\parallel$$\bm y$) and $\bm E^{\omega}$$\parallel$[110] ($\parallel$$\bm z$) under $\bm H_{\rm dc}$$\parallel$[110]. In the calculation, an isotropic dielectric tensor with $\epsilon_{zz}^{\infty}$=$\epsilon_{yy}^{\infty}$=$\epsilon^{\infty}$=8 is used according to the dielectric-measurement data~\cite{Belesi12,Miller10}, while $\mu_{zz}^{\infty}$=$\mu_{yy}^{\infty}$=1 is used for permeability. The value of $J$ is set to be $J$=1 meV. 

Note that the value of exchange parameter in Cu$_2$OSeO$_3$ is $J$$\sim$3 meV in reality, but here we used $J$=1 meV to reproduce ciritcal fields at a finite temperature in order to compare the calculation and the experiment done at $T$=57.5 K. For an exact numerical treatment, we should perform a finite-temperature calculation by taking account of thermal-fluctuation effects using the stochastic Landau-Lifshitz-Gilbert equation with a realistic exchange parameter of $J$=3 meV. The simple reduction of the $J$ value adopted here to mimic the thermal effects is a rather crude approximation. As a result, the calculation tends to overestimate the magnitude of directional dichroism at finite temperatures significantly. However, even with this approximation, not only qualitative behaviors of physical phenomena but also experimentally measured resonance frequnecies and magnetic-field strengths can be reproduced quantitatively. Also, the predicted large microwave directional dichroism of $\sim$20-30 $\%$ is expected to be observed at low temperatures.

\subsection{Nonreciprocal directional dichroism: Experiment}
\begin{figure*}
\begin{center}
\includegraphics[width=2.0\columnwidth]{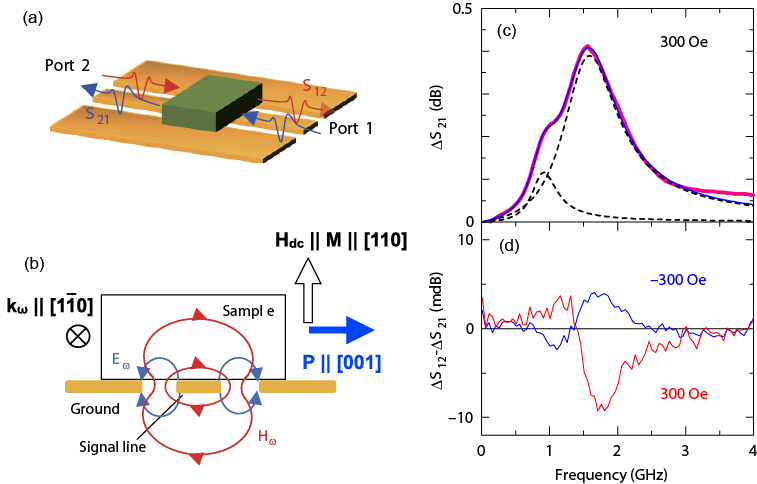}
\end{center}
\caption{(color online). (a) Experimental setup of the measurement of microwave directional dichroism. Microwaves can propagate along the coplanar waveguide in both directions, and the difference between $\Delta S_{21}$ and $\Delta S_{12}$ is evaluated as nonreciprocal absorption spectra. (b) Electromagnetic-field distribution in the present setup using a coplanar waveguide, viewed from the direction of microwave propagation. Here the configuration of $\bm P \parallel [001]$, $\bm H_{\rm dc} \parallel \bm M \parallel [110]$, and $\bm K^\omega \parallel [1\bar{1}0]$ is taken for bulk Cu$_2$OSeO$_3$, and thus the relationship $\bm P \times \bm M \parallel \bm K^\omega$ is satisfied. (c) Microwave absorption spectra for the skyrmion-crystal state at $H_{\rm dc} = 300$ Oe. A result of the two-peak fitting is also shown by broken lines in addition to the raw data (red line). The broken lines represent each component, and the blue solid line represents a sum of them. (d) The corresponding nonreciprocal absorption spectra measured for $H_{\rm dc}=\pm 300$ Oe. (Reproduced from Ref.~\cite{Okamura13}.)}
\label{Fig24}
\end{figure*}
\begin{figure*}
\begin{center}
\includegraphics[width=2.0\columnwidth]{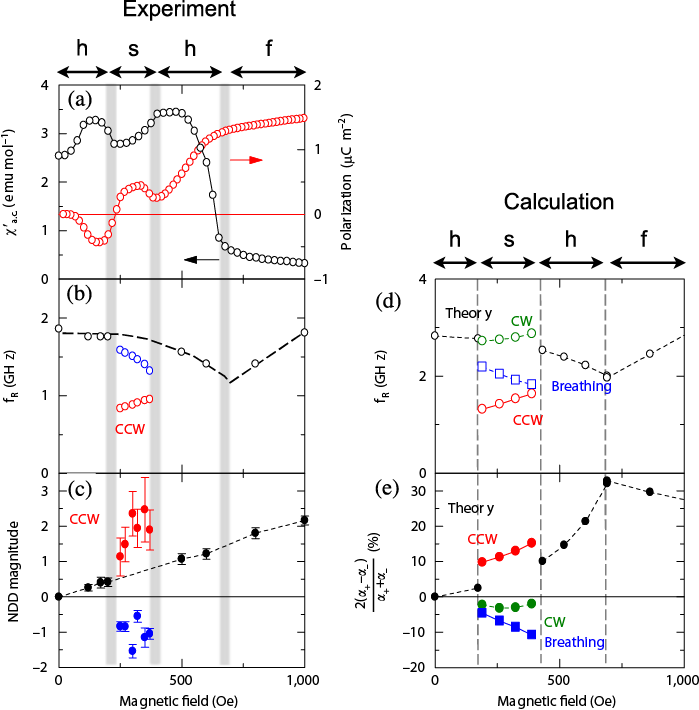}
\end{center}
\caption{(color online). Comparison between experiment and theory of the skyrmion magnetoelectric resonance. (a)-(c) $H_{\rm dc}$ dependence of (a) ac magnetic susceptibility for $\bm H_{\rm dc} \parallel [110]$, (b) resonance frequency $f_{\rm R}$, and (c) normalized magnitude of microwave directional dichroism. In (a), the electric polarization profile is also presented. Broken line in (b) is a guide to the eyes. (d) and (e) Calculated magnetic-field dependence of (d) the resonance frequencies $f_{\rm R}$ and (e) the magnitudes of the directional dichroism normalized by the absorption coefficient for $T$=0. The thick vertical lines in (a)-(c) and the broken vertical lines in (d) and (e) represent the boundaries between different magnetic phases: from low to high fields, the magnetic phases are helical, skyrmion-crystal, conical and collinear phases. (Reproduced from Ref.~\cite{Okamura13}.)}
\label{Fig25}
\end{figure*}
The above theoretical analysis suggests that significant directional dichroism can be expected for the skyrmion resonant modes in Cu$_2$OSeO$_3$ under the experimental configuration with $\bm H_{\rm dc}$$\parallel$[110] (eventually $\bm P$$\parallel$[001]) and $\bm K^\omega$$\parallel$$[1\bar{1}0]$ since the condition $\bm{P} \times \bm{M} \parallel \bm K^\omega$ is satisfied here. Figure~\ref{Fig24}(b) shows the nonreciprocal absorption spectra $\Delta\alpha(\omega)=\alpha_{+}(\omega) - \alpha_{-}(\omega)$ measured for the skyrmion-crystal phase in bulk Cu$_2$OSeO$_3$ at $T$=57 K~\cite{Okamura13}. Two resonant modes observed at 1.0 GHz and 1.7 GHz correspond to the counterclockwise rotational mode and the breathing mode of the skyrmion crystal, respectively. Note that for the experimental setup [see Fig.~\ref{Fig24}(a)] with electromagnetic-field distribution shown in Fig.~\ref{Fig24}(b), both the rotational mode active to $\bm H^\omega$$\perp$$\bm H_{\rm dc}$ and the breathing mode active to $\bm H^\omega$$\parallel$$\bm H_{\rm dc}$ are simultaneously excited [see Fig.~\ref{Fig24}(c)]. Both modes show large directional dichroism up to $\sim3\%$ but with opposite signs [see Fig.~\ref{Fig24}(d)]. Here the reversal of $H_{\rm dc}$, which reverses the sign of $\bm M$ but does not change the sign of $\bm P$, changes the sign of directional dichroism, which is consistent with the symmetry requirement of this phenomena. The experimentally and theoretically obtained $H_{\rm dc}$-dependence of the magnetic resonance frequency as well as the magnitude of directional dichroism for bulk Cu$_2$OSeO$_3$ are summarized in Fig.~\ref{Fig25}, which agree well with each other~\cite{Okamura13}. The experimental observation of the large directional dichroism inversely proves that the present skyrmion resonant modes have finite electric activity coupled with $\bm E^\omega$. The above results demonstrate that ultra-fast control of skyrmions by ac electric fields up to GHz frequency range is indeed possible in insulating materials, and the employment of resonant structures in the electric susceptibility as presented here may be useful to improve the efficiency of the electric manipulation of magnetic skyrmions.

\subsection{Microwave magnetochiral effect}
\begin{figure}
\begin{center}
\includegraphics[width=1.0\columnwidth]{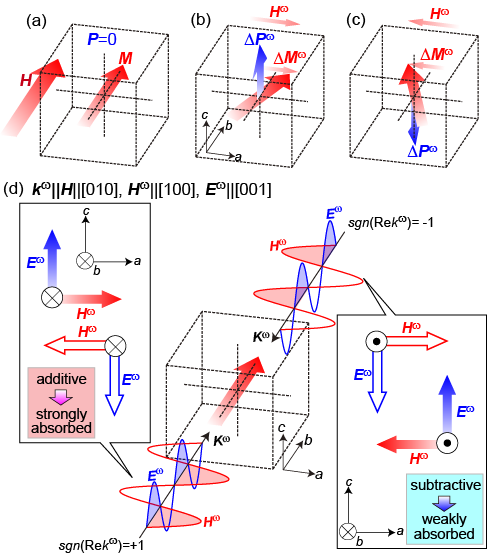}
\end{center}
\caption{(color online). (a)-(c) In the presence of net magnetization $\bm M$$\parallel$$\bm H$ under $\bm H$$\parallel$[010], oscillating magnetization component $\Delta \bm M^\omega$($\parallel$[100]) is accompanied by the oscillating polarization component $\Delta \bm P^\omega$($\parallel$[001]). (d) Configuration of microwave $\bm H^\omega$ and $\bm E^\omega$ components, for which the magnetochiral dichroism occurs under $\bm H$$\parallel$[010]: $\bm K^\omega$$\parallel$$\pm \bm H$, $\bm H^\omega$$\parallel$[100] and $\bm E^\omega$$\parallel$[001]. (Reproduced from Ref.~\cite{Mochizuki15}.)}
\label{Fig26}
\end{figure}
In addition to the Voigt geometry with $\bm K^\omega$$\perp$$\bm H_{\rm dc}$ as discussed above, one can expect the microwave directional dichroism also for the Faraday geometry with $\bm K^\omega$$\parallel$$\bm H_{\rm dc}$~\cite{Okamura15,Mochizuki15}. The directional dichroism realized with this configuration is called magnetochiral effect where the absorption or transmission intensity of an electromagnetic wave propagating parallel or antiparallel to the external magnetic field $\bm H_{\rm dc}$ differs depending on its propagation direction.

Considering the emergence of finite $\bm P$$\parallel$[001]($\parallel$$\bm c$) under $\bm H_{\rm dc}$$\parallel$[110] and the absence of $\bm P$ ($\bm P$=0) under $\bm H_{\rm dc}$$\parallel$[010]($\parallel$$\bm b$) as shown, respectively, in Fig.~\ref{Fig09}(b) and Fig.~\ref{Fig26}(a), one expects the emergence of oscillating ferroelectric polarization $\Delta \bm P^\omega$$\parallel$[001]($\parallel$$\bm c$) induced by the oscillation of net magnetization $\bm M$ with $\Delta \bm M^\omega$$\parallel$[100]($\parallel$$\bm a$) when the ac magnetic field $\bm H^\omega$$\parallel$[100]($\parallel$$\bm a$) is applied to the Cu$_2$OSeO$_3$ sample under $\bm H_{\rm dc}$$\parallel$[010]($\parallel$$\bm b$) [see Figs.~\ref{Fig26}(b) and (c)].

Consequently the interference between the $\bm H^\omega$-active and the $\bm E^\omega$-active processes of this coupled oscillation of $\Delta \bm M^\omega$ and $\Delta \bm P^\omega$, and thereby the directional dichroism occur for microwave configuration of $\bm K^\omega$$\parallel$[010]($\parallel$$\bm b$), $\bm H^\omega$$\parallel$[100]($\parallel$$\bm a$) and $\bm E^\omega$$\parallel$[001]($\parallel$$\bm c$) as shown in Fig.~\ref{Fig26}(d). Namely the $\bm H^\omega$ and $\bm E^\omega$ components of an microwave propagating in [010] or $+\bm b$ direction cooperatively excite the above-mentioned electromagnon mode and hence is absorbed significantly, while those of an oppositely propagating microwave destructively contribute to the electromagnon excitation and hence is absorbed only weakly.

The microwave magnetochiral effect with this configuration is expected not only for the skyrmion-crystal phase but also for other magnetically ordered phases. The effect in the conical phase and the field-polarized ferromagnetic phase in the bulk Cu$_2$OSeO$_3$ sample at low temperatures was experimentally observed~\cite{Okamura15} and theoretically predicted~\cite{Mochizuki15} individually around the same time. 

\section{Summary and Perspectives}
In this article, we have overviewed recent theoretical and experimental studies on the multiferroic properties and the dynamical magnetoelectric phenomena of magnetic skyrmions in a chiral-lattice magnetic insulator Cu$_2$OSeO$_3$. We have first discussed that the noncollinear skyrmion spin textures in this insulating magnet induce electric polarizations via the so-called spin-dependent metal-ligand hybridization mechanism, and the system attains multiferroic nature. Resulting magnetoelectric coupling, that is, the coupling between magnetizations and polarizations enables us to manipulate magnetic skyrmions by application of electric fields instead of injection of electric currents. Here, an important future issue is an establishment of a method to create skyrmions by electric fields. This method will be microscopically distinct from that based on spin-transfer torques from spin-polarized electric currents in metallic systems, and can be a unique technique for future skyrmion-based storage devices without energy losses due to Joule heating.

We have then argued that multiferroic skyrmions show coupled oscillation modes of magnetizations and polarizations, so-called electromagnon excitations, which are both magnetically and electrically active. Interference between these electric and magnetic activation processes leads to peculiar magnetoelectric effects of skyrmions in the microwave frequency regime. Significant dynamical magnetoelectric effects such as directional dichroism are rare for any frequency ranges. In particular there have been few reports on their observations in the gigahertz regime. This is because usual multiferroic materials based on simple short-period spin structures with antiferromagnetic interactions tend to have rather high resonance frequencies (typically at the terahertz regime) due to large spin-wave gaps. In turn, skyrmions and other topological spin textures with long-period spin modulations induced by the Dzyaloshinskii--Moriya interactions and ferromagnetic interactions tend to have small spin-wave gaps and specific low-lying resonant modes, which offers a unique opportunity to realize novel microwave functionalities. 

For technical applications of these effects to future microwave devices~\cite{Gurevich96}, it is important to achieve enhanced cross-correlation responses or larger ferroelectric polarization. For this purpose, search for new insulating materials which exhibit skyrmions is necessary~\cite{KezsmarkiCD}. Materials containing ions with stronger spin-orbit interactions other than Cu$^{2+}$ is one of the promising research directions. In addition, it is also promising to explore skyrmion states which induce large ferroelectric polarizations via a mechanism other than the spin-dependent hybridization mechanism such as the inverse Dzyaloshinskii-Moriya mechanism~\cite{Katsura05}.

The insulating skyrmionic material also offers a unique opportunity to investigate skyrmion motion driven by magnon currents. It has been theoretically predicted that magnon currents can drive translational motion, subsequent Hall motion, and rotational motion of skyrmions via the spin-transfer torques~\cite{Mochizuki14,KongL13,LinSZ14a,Iwasaki14,Schutte14}. In particular, thermally-induced diffusive flows of magnons are expected to induce the skyrmion motion in the presence of temperature gradient~\cite{Mochizuki14,KongL13,LinSZ14a}. In the metallic magnets, a thermal gradient necessarily induces diffusive currents of conduction electrons which flow from a higher-temperature side to a lower-temperature side, and contribute to the skyrmion motion. Therefore it is difficult to identify a contribution purely from magnon currents to the skyrmion motion in metallic magnets. In turn, insulating magnets are appropriate for the research because of the absence of conduction electrons and low-lying charge excitations. The skyrmion-hosting insulating magnets offer precious playgrounds for research into several interesting issues such as optical responses and manipulations of skyrmions~\cite{Finazzi13,Koshibae14,Ogawa15}, skyrmion-based magnonic crystals~\cite{Krawczyk14,MaF15}, spin currents~\cite{Hirobe15}, spin motive forces~\cite{Shimada15}, spin Seebeck effects~\cite{LinSZ14a,Kovalev14} in the skyrmion phases, elastic responses of skyrmions~\cite{Nii14}, $E$-field driven motion and manipulations of skyrmions~\cite{White12,White14,LiuYH13}, and critical and dynamical behaviors of phase transitions~\cite{Janoschek13,Zivkovic14,LinSZ14c,Levatic14}.

So far, magnetic skyrmions had been observed only in materials with chiral cubic P$_{2_13}$ symmetry such as B20 alloys and Cu$_2$OSeO$_3$. However, it was theoretically predicted that non-chiral but polar magnets with C$_{nv}$ symmetry can also host skyrmions stabilized by the Dzyaloshinskii--Moriya interactions~\cite{Bogdanov89}. Indeed, realization of magnetic skyrmions was recently discovered in polar magnet GaV$_4$S$_8$ with rhombohedral C$_{3v}$ symmetry~\cite{KezsmarkiCD}. In addition to these non-centrosymmetric ferromagnets, observations of skyrmions have been reported even for centrosymmetric ferromagnetic insulators with uniaxial anisotropy such as Y$_3$Fe$_5$O$_{12}$~\cite{LinYS73,Giess80}, $R$FeO$_3$~\cite{LinYS73,Giess80}, BaFe$_{11.79}$Sc$_{0.16}$Mg$_{0.05}$O$_{19}$~\cite{YuXZ12b}, La$_{0.5}$Ba$_{0.5}$MnO$_3$~\cite{Nagao13}, La$_{2-2x}$Sr$_{1+2x}$Mn$_2$O$_7$~\cite{YuXZ14}. Crystal structures of these materials have spatial inversion symmetry, and thus the Dzyaloshinskii--Moriya interaction is not active. Instead, interplay between magnetic dipole--dipole interaction and magnetic anisotropies play important roles for the realization of magnetic skyrmions. Furthermore, it was recently discovered that surfaces and interfaces of ferromagnetic monolayers host atomically small magnetic skyrmions~\cite{Heinze00,Heinze11,Romming13}. There, the space-inversion symmetry is broken, and thus the Dzyaloshinskii--Moriya interaction is active. Now the number of known skyrmion-hosting magnetic systems is rapidly increasing. Magnetoelectric dynamics in these systems are issues of interest and should be clarified in future studies.

\section{Appendix}
\label{sec06}
\noindent
{\bf Derivation of the expression of $N(\omega)$}\\

In this Appendix, we solve Maxwell's equations to derive the expression of $N(\omega)$ given by Eq.~(\ref{eqn:CRIHyEz}) for the following polarization configuration of electromagnetic wave as an example:
$\bm H^{\omega}(\parallel \bm y)=(0, H_y^{\omega}, 0)$, 
$\bm E^{\omega}(\parallel \bm z)=(0, 0, E_z^{\omega})$, and
$\bm K^{\omega}(\parallel \bm x)=(K^{\omega}, 0, 0)$. 
\\
Inserting Eq.~(\ref{eqn:InducedB}) into Eq.~(\ref{eqn:MaxwellEq1}), we obtain
\begin{eqnarray}
\omega \mu_0 \hat{\tilde{\chi}}^{\rm mm}{\bm H}^\omega 
+\omega \sqrt{\epsilon_0 \mu_0} \hat{\chi}^{\rm me} {\bm E}^\omega
&=&\bm K^{\omega} \times \bm E^{\omega}
\nonumber \\
&=&(0, -K^{\omega}E_z^{\omega}, 0),
\end{eqnarray}
where $\hat{\tilde{\chi}}^{\rm mm}=\hat{\mu}^\infty +\hat{\chi}^{\rm mm}$.
\\
Also inserting Eq.~(\ref{eqn:InducedD}) into Eq.~(\ref{eqn:MaxwellEq2}), we obtain
\begin{eqnarray}
-\omega \epsilon_0 \hat{\tilde{\chi}}^{\rm ee}{\bm E}^\omega 
-\omega \sqrt{\epsilon_0 \mu_0} \hat{\chi}^{\rm em} {\bm H}^\omega
&=&\bm K^{\omega} \times \bm H^{\omega}
\nonumber \\
&=&(0, 0, -K^{\omega}H_y^{\omega}),
\end{eqnarray}
where $\hat{\tilde{\chi}}^{\rm ee}=\hat{\epsilon}^\infty +\hat{\chi}^{\rm ee}$.\\
Now we have a set of equations,
\begin{eqnarray}
\omega \mu_0 H_y^\omega \tilde{\chi}^{\rm mm}_{yy}
+ \omega \sqrt{\epsilon_0 \mu_0} E_z^\omega \chi^{\rm me}_{yz}
=-K^{\omega} E_z^\omega, \\
\nonumber \\
\omega \epsilon_0 E_z^\omega \tilde{\chi}^{\rm ee}_{zz}
+ \omega \sqrt{\epsilon_0 \mu_0} H_y^\omega \chi^{\rm em}_{zy}
=-K^{\omega} H_y^\omega.
\end{eqnarray}
After multiplying $\frac{c}{\omega}=\frac{1}{\omega\sqrt{\epsilon_0 \mu_0}}$ to both sides of the equations and using $N=\frac{cK^{\omega}}{\omega}$, the equations lead,
\begin{eqnarray}
\left(
\begin{array}{cc}
\chi^{\rm me}_{yz}+{\rm sgn}(K^{\omega})N & 
\sqrt{\frac{\mu_0}{\epsilon_0}}\tilde{\chi}^{\rm mm}_{yy}  \\
\sqrt{\frac{\epsilon_0}{\mu_0}}\tilde{\chi}^{\rm ee}_{zz} & 
\chi^{\rm em}_{zy}+{\rm sgn}(K^{\omega})N
\end{array}
\right) \left(
\begin{array}{cc}
E_z^\omega\\  H_y^\omega
\end{array}
\right)=0. \nonumber \\
\end{eqnarray}
These equations have nontrivial solutions when the determinant of the matrix is zero, that is,
\begin{eqnarray}
\chi^{\rm me}_{yz} \chi^{\rm em}_{zy}
+{\rm sgn}(K^{\omega})N(\chi^{\rm me}_{yz} + \chi^{\rm em}_{zy})
+N^2 - \tilde{\chi}^{\rm ee}_{zz}\tilde{\chi}^{\rm mm}_{yy}=0.
\nonumber \\
\end{eqnarray}
Neglecting the first term of the left-hand side because $\chi^{\rm me}_{yz}$ and $\chi^{\rm em}_{zy}$ are small, the equation leads
\begin{eqnarray*}
N^2+{\rm sgn}(K^{\omega})N(\chi^{\rm me}_{yz} + \chi^{\rm em}_{zy})
-\tilde{\chi}^{\rm ee}_{zz}\tilde{\chi}^{\rm mm}_{yy}=0.
\end{eqnarray*}
Solving this quadratic equation, we obtain
\begin{eqnarray}
N=\frac
{-{\rm sgn}(K^{\omega})(\chi^{\rm me}_{yz} + \chi^{\rm em}_{zy})
\pm \sqrt{(\chi^{\rm me}_{yz} + \chi^{\rm em}_{zy})^2
+4\tilde{\chi}^{\rm ee}_{zz}\tilde{\chi}^{\rm mm}_{yy}}}
{2}.
\nonumber \\
\end{eqnarray}
After neglecting the second-order terms with respect to $\chi^{\rm me}_{yz}$ and $\chi^{\rm em}_{zy}$ again, we eventually arrive at
\begin{eqnarray}
N(\omega)=\sqrt{\tilde{\chi}^{\rm ee}_{zz}\tilde{\chi}^{\rm mm}_{yy}}
-{\rm sgn}(K^{\omega})(\chi^{\rm me}_{yz} + \chi^{\rm em}_{zy})/2.
\end{eqnarray}

\section{Acknowledgement}
The authors thank Y. Tokura, N. Nagaosa, Y. Onose, F. Kagawa, X. Z. Yu, Y. Okamura, M. Kawasaki, and A. Rosch for enlightening discussions. This work is partly supported by JSPS KAKENHI (Grant Numbers 25870169, 25287088, 26610109 and 15H05458) and JST PRESTO.
\\


\begin{thebibliography}{99.}%
\bibitem{Fiebig05}M. Fiebig, J. Phys. D: Appl. Phys. {\bf 38}, R123 (2005).
\bibitem{Tokura06}Y. Tokura, Science {\bf 312}, 1481 (2006).
\bibitem{Khomskii06}D. I. Khomskii, J. Magn. Magn. Mater. {\bf 306}, 1 (2006).
\bibitem{Cheong07}S.-W. Cheong and M. Mostovoy, Nature Mater. {\bf 6}, 13 (2007).
\bibitem{Tokura07}Y. Tokura, J. Magn. Magn. Mater. {\bf 310}, 1145 (2007).
\bibitem{Tokura14}Y. Tokura, S. Seki, N. Nagaosa, Rep. Prog. Phys. {\bf 77}, 076501 (2014).
\bibitem{Katsura05}H. Katsura, N. Nagaosa, and A. V. Balatsky, Phys. Rev. Lett. {\bf 95}, 057205 (2005).

\bibitem{Mostovoy06}M. Mostovoy, Phys. Rev. Lett. {\bf 96}, 067601 (2006).
\bibitem{Kimura03}T. Kimura, T. Goto, H. Shintani, K. Ishizaka, T. Arima, and Y. Tokura, Nature {\bf 426}, 55 (2003).

\bibitem{Yamasaki07}Y. Yamasaki, H. Sagayama, T. Goto, M. Matsuura, K. Hirota, T. Arima, and Y. Tokura, Phys. Rev. Lett. {\bf 98}, 147204 (2007).
\bibitem{Seki08}S. Seki, Y. Yamasaki, M. Soda, M. Matsuura, K. Hirota, and Y. Tokura, Phys. Rev. Lett. {\bf 100}, 127201 (2008).
\bibitem{Murakawa09}H. Murakawa, Y. Onose, Y. Tokura, Phys. Rev. Lett. {\bf 103}, 147201 (2009).
\bibitem{Tokunaga09}Y. Tokunaga, N. Furukawa, H. Sakai, Y. Taguchi, T. Arima, and Y. Tokura, Nature Mater. {\bf 8}, 558 (2009).
\bibitem{Pfleiderer11}C. Pfleiderer, Nature Phys. {\bf 7}, 673 (2011).

\bibitem{Nagaosa13}N. Nagaosa, and Y. Tokura, Nature Nanotech. {\bf 8}, 899 (2013).

\bibitem{Fert13}A. Fert, V. Cros, and J. Sampaio, Nature Nanotech. {\bf 8}, 152 (2013).
\bibitem{Skyrme61}T. H. R. Skyrme, Proc. R. Soc. A {\bf 260}, 127 (1961).

\bibitem{Skyrme62}T. H. R. Skyrme, Nucl. Phys. {\bf 31}, 556 (1962).
\bibitem{Bogdanov89}A.N. Bogdanov, and D.A. Yablonskii, Sov. Phys. JETP {\bf 68}, 101 (1989).

\bibitem{Bogdanov94}A. Bogdanov, and A. Hubert, J. Mag. Mag. Mat. {\bf 138}, 255 (1994).

\bibitem{Rossler06}U.K. R\"o{\ss}ler, A. N. Bogdanov, and C. Pfleiderer, Nature {\bf 442}, 797 (2006).
\bibitem{Dzyaloshinskii58}I. Dzyaloshinskii, J. Phys. Chem. Solids {\bf 4}, 241 (1958).

\bibitem{Moriya60}T. Moriya, Phys. Rev. {\bf 120}, 91 (1960).
\bibitem{Muhlbauer09}S. M\"uhlbauer, B. Binz, F. Jonietz, C. Pfleiderer, A. Rosch, A. Neubauer, R. Georgii, and P. B\"oni, Science {\bf 323}, 915 (2009).

\bibitem{YuXZ10}X. Z. Yu, Y. Onose, N. Kanazawa, J. H. Park, J. H. Han, Y. Matsui, N. Nagaosa, and Y. Tokura, Nature {\bf 465}, 901 (2010).
\bibitem{Pappas09}C. Pappas, E. Lelie\`vre-Berna, P. Falus, P. M. Bentley, E. Moskvin, S. Grigoriev, P. Fouquet, and B. Farago, Phys. Rev. Lett. {\bf 102}, 197202 (2009).

\bibitem{Pfleiderer10}C. Pfleiderer, T. Adams, A. Bauer, W. Biberacher, B. Binz, F. Birkelbach, P. B\"oni, C. Franz, R. Georgii, M. Janoschek, F. Jonietz, T. Keller, R. Ritz, S. M\"uhlbauer, W. Munzer, A. Neubauer, B. Pedersen, and A. Rosch, J. Phys. Condens. Matter {\bf 22}, 164207 (2010).

\bibitem{Munzer10}W. Munzer, A. Neubauer, T. Adams, S. M\"uhlbauer, C. Franz, F. Jonietz, R. Georgii, P. B\"oni, B. Pedersen, M. Schmidt, A. Rosch, and C. Pfleiderer, Phys. Rev. B {\bf 81}, 041203(R) (2010).

\bibitem{Adams11}T. Adams, S. M\"uhlbauer, C. Pfleiderer, F. Jonietz, A. Bauer, A. Neubauer, R. Georgii, P. B\"oni, U. Keiderling, K. Everschor, M. Garst, and A. Rosch, Phys. Rev. Lett. {\bf 107}, 217206 (2011).

\bibitem{Grigoriev13}S. V. Grigoriev, N. M. Potapova, S.-A. Siegfried, V. A. Dyadkin, E. V. Moskvin, V. Dmitriev, D. Menzel, C. D. Dewhurst, D. Chernyshov, R. A. Sadykov, L. N. Fomicheva, and A. V. Tsvyashchenko, Phys. Rev. Lett. {\bf 110}, 207201 (2013).
\bibitem{YuXZ11}X. Z. Yu, N. Kanazawa, Y. Onose, K. Kimoto, W. Z. Zhang, S. Ishiwata, Y. Matsui, and Y. Tokura, Nature Mater. {\bf 10}, 106 (2011).

\bibitem{Tonomura12}A. Tonomura X. Z. Yu, K. Yanagisawa, T. Matsuda, Y. Onose, N. Kanazawa, H. S. Park, and Y. Tokura, Nano Lett. {\bf 12}, 1673 (2012).

\bibitem{Shibata13}K. Shibata, X. Z. Yu, T. Hara, D. Morikawa, N. Kanazawa, K. Kimoto, S. Ishiwata, Y. Matsui, and Y. Tokura, Nature Nanotech. {\bf 8}, 723 (2013).

\bibitem{Morikawa13}D. Morikawa, K. Shibata, N. Kanazawa, X. Z. Yu, and Y. Tokura, Phys. Rev. B {\bf 88}, 024408 (2013).
\bibitem{YiSD09}S. D. Yi, S. Onoda, N. Nagaosa, and J. H. Han, Phys. Rev. B {\bf 80}, 054416 (2009).

\bibitem{HanJH10}J. H. Han, J. Zang, Z. Yang, J.-H. Park, and N. Nagaosa, Phys. Rev. B {\bf 82}, 094429 (2010).

\bibitem{LiYQ11}Y.-Q. Li, Y.-H. Liu, and Y. Zhou, Phys. Rev. B {\bf 84}, 205123 (2011).
\bibitem{Butenko10}A. B. Butenko, A. A. Leonov, U. K. R\"o{\ss}ler, and A. N. Bogdanov, Phys. Rev. B {\bf 82}, 052403 (2010).

\bibitem{Kiselev11}N. S. Kiselev, A. N. Bogdanov, R. Sch\"afer, and U. K. R\"o{\ss}ler, J. Phys. D {\bf 44}, 392001 (2011).

\bibitem{Wilson12}M. N. Wilson, E. A. Karhu, A. S. Quigley, U. K. R\"o{\ss}ler, A. B. Butenko, A. N. Bogdanov, M. D. Robertson, and T. L. Monchesky, Phys. Rev. B {\bf 86}, 144420 (2012).

\bibitem{Karhu12}E. A. Karhu, U. K. R\"o{\ss}ler, A. N. Bogdanov, S. Kahwaji, B. J. Kirby, H. Fritzsche, M. D. Robertson, C. F. Majkrzak, and T. L. Monchesky, Phys. Rev. B {\bf 85}, 094429 (2012).

\bibitem{Rybakov13}F. N. Rybakov, A. B. Borisov, and A. N. Bogdanov, Phys. Rev. B {\bf 87}, 094424 (2013).

\bibitem{Kwon12}H. Y. Kwon, K. M. Bu, Y. Z. Wu, and C. Won, J. Magn. Magn. Mater. 324, 2171 (2012).

\bibitem{Jonietz10}F. Jonietz, S. M\"uhlbauer, C. Pfleiderer, A. Neubauer, W. M\"unzer, A. Bauer, T. Adams, R. Georgii, P. B\"oni, R. A. Duine, K. Everschor, M. Garst, and A. Rosch, Science {\bf 330}, 1648 (2010).

\bibitem{YuXZ12}X. Z. Yu, N. Kanazawa, W.Z. Zhang, T. Nagai, T. Hara, K. Kimoto, Y. Matsui, Y. Onose, and Y. Tokura, Nature Commun. {\bf 3}, 988 (2012).

\bibitem{Everschor11}K. Everschor, M. Garst, R. A. Duine, and A. Rosch, Phys. Rev. B {\bf 84}, 064401 (2011).

\bibitem{Everschor12}K. Everschor, M. Garst, B. Binz, F. Jonietz, S. M\"uhlbauer, C. Pfleiderer, and A. Rosch, Phys. Rev. B {\bf 86}, 054432 (2012).

\bibitem{Iwasaki13a}J. Iwasaki, M. Mochizuki, and N. Nagaosa, Nature Commun. {\bf 4}, 1463 (2013).

\bibitem{Iwasaki13b}J. Iwasaki, M. Mochizuki, and N. Nagaosa, Nature Nanotech. {\bf 8}, 742 (2013).

\bibitem{Rosch13}A. Rosch, Nature Nanotech. {\bf 8}, 160 (2013).

\bibitem{Nagaosa12}N. Nagaosa, and Y. Tokura, Phys. Scr. T {\bf 146}, 014020 (2012).

\bibitem{LeeM07}M. Lee, Y. Onose, Y. Tokura, and N. P. Ong, Phys. Rev. B {\bf 75}, 172403 (2007).

\bibitem{Binz08}B. Binz, and A. Vishwanath, Physica B {\bf 403}, 1336 (2008).

\bibitem{Neubauer09}A. Neubauer, C. Pfleiderer, B. Binz, A. Rosch, R. Ritz, P. G. Niklowitz, and P. B\"oni, Phys. Rev. Lett. {\bf 102}, 186602 (2009).

\bibitem{ZangJ11}J. Zang, M. Mostovoy, J. H. Han, and N. Nagaosa, Phys. Rev. Lett. {\bf 107}, 136804 (2011).

\bibitem{Kanazawa11}N. Kanazawa, Y. Onose, T. Arima, D. Okuyama, K. Ohoyama, S. Wakimoto, K. Kakurai, S. Ishiwata, and Y. Tokura, Phys. Rev. Lett. {\bf 106}, 156603 (2011).

\bibitem{Schulz12}T. Schulz, R. Ritz, A. Bauer, M. Halder, M. Wagner, C. Franz, C. Pfleiderer, K. Everschor, M. Garst, and A. Rosch, Nature Phys. {\bf 8}, 301 (2012).

\bibitem{Milde13}P. Milde, D. K\"ohler, J. Seidel, L. M. Eng, A. Bauer, A. Chacon, J. Kindervater, S. M\"uhlbauer, C. Pfleiderer, S. Buhrandt, C. Sch\"ute, and A. Rosch, Science {\bf 340}, 1076 (2013).

\bibitem{Takashima14}R. Takashima, and S. Fujimoto, J. Phys. Soc. Jpn. {\bf 83}, 054717 (2014).

\bibitem{Seki12a}S. Seki, X. Z. Yu, S. Ishiwata, and Y. Tokura, Science {\bf 336}, 198 (2012).

\bibitem{Seki12b}S. Seki, J.-H. Kim, D. S. Inosov, R. Georgii, B. Keimer, S. Ishiwata, and Y. Tokura, Phys. Rev. B {\bf 85}, 220406 (2012).

\bibitem{Adams12}T. Adams, A. Chacon, M. Wagner, A. Bauer, G. Brandl, B. Pedersen, H. Berger, P. Lemmens, and C. Pfleiderer, Phys. Rev. Lett. {\bf 108}, 237204 (2012).

\bibitem{Bos08}J.-W. G. Bos, C. V. Colin, and T. T. M. Palstra, Phys. Rev. B {\bf 78}, 094416 (2008).

\bibitem{Belesi10}M. Belesi, I. Rousochatzakis, H. C. Wu, H. Berger, I. V. Shvets, F. Mila, and J. P. Ansermet, Phys. Rev. B {\bf 82}, 094422 (2010).

\bibitem{Belesi11}M. Belesi, T. Philippe, I. Rousochatzakis, H. C. Wu, H. Berger, S. Granville, I. V. Shvets, and J.-Ph. Ansermet, J. Phys.: Conf. Ser. {\bf 303}, 012069 (2011).

\bibitem{Curie1894} P. Curie, J. Physique {\bf 3}, 393 (1894).

\bibitem{Kimura07}T. Kimura, Annu. Rev. Mater. Res. {\bf 37}, 387 (2007).

\bibitem{Seki12c}S. Seki, S. Ishiwata, and Y. Tokura, Phys. Rev. B {\bf 86}, 060403 (2012).

\bibitem{Maisuradze12}A. Maisuradze, A. Shengelaya, H. Berger, D. M. Djoki\'c, and H. Keller, Phys. Rev. Lett. {\bf 108}, 247211 (2012).

\bibitem{Omrani14}A. A. Omrani, J. S. White, K. Pr{\v s}a, I. {\v Z}ivkovi\'c, H. Berger, A. Magrez, Ye-Hua Liu, J. H. Han, and H. M. Ronnow, Phys. Rev. B {\bf 89}, 064406 (2014).

\bibitem{Ruff15}E. Ruff, P. Lunkenheimer, A. Loidl, H. Berger, S. Krohns, Sci. Rep. {\bf 5}, 15025 (2015).

\bibitem{Langner14}M.-C. Langner, S. Roy, S.-K. Mishra, J.-C.-T. Lee, X.-W. Shi, M.-A. Hossain, Y.-D. Chuang, S. Seki, Y. Tokura, S.-D. Kevan, and R.-W. Schoenlein, Phys. Rev. Lett. {\bf 112}, 167202 (2014).

\bibitem{Lancaster15}T. Lancaster, R. C. Williams, I. O. Thomas, F. Xiao, F. L. Pratt, S. J. Blundell, J. C. Loudon, T. Hesjedal, S. J. Clark, P. D. Hatton, M. C. Hatnean, D. S. Keeble, and G. Balakrishnan, Phys. Rev. B {\bf 91}, 224408 (2015).


\bibitem{Yang12}J. H. Yang, Z. L. Li, X. Z. Lu, M.-H. Whangbo, S.-H. Wei, X. G. Gong, H. J. Xiang, Phys. Rev. Lett. {\bf 109}, 107203 (2012).

\bibitem{Jia06}C. Jia, S. Onoda, N. Nagaosa, J. H. Han, Phys. Rev. B {\bf 74}, 224444 (2006).

\bibitem{Jia07}C. Jia, S. Onoda, N. Nagaosa, J. H. Han, Phys. Rev. B {\bf 76}, 144424 (2007).

\bibitem{Arima07}T. Arima, J. Phys. Soc. Jpn. {\bf 76}, 073702 (2007).

\bibitem{White12}J. S. White, I. Levati\'{c}, A. A. Omrani, N. Egetenmeyer, K. Prsa, I. Zivkovic, J. L. Gavilano, J. Kohlbrecher, M. Bartkowiak, H. Berger, H. M. R{\o}nnow, J. Phys.: Condens. Matter {\bf 24}, 432201 (2012).

\bibitem{White14}J. S. White, P. Prsa, P. Huang, A. A. Omrani, I. Zivkovic, M. Bartkowiak, H. Berger, A. Magrez, J. L. Gavilano, G. Nagy, J. Zang, H. M. R{\o}nnow, Phys. Rev. Lett. {\bf 113}, 107203 (2014).

\bibitem{Mochizuki13}M. Mochizuki, and S. Seki, Phys. Rev. B {\bf 87}, 134403 (2013).

\bibitem{Liu13}Y. H. Liu, Y.-Q. Li, and J. H. Hoon, Phys. Rev. B {\bf 87}, 100402 (2013).

\bibitem{Bak80}P. Bak, and M. H. Jensen, J. Phys. C {\bf 13}, L881 (1980).

\bibitem{Buhrandt13}S. Buhrandt, and L. Fritz, Phys. Rev. B {\bf 88}, 195137 (2013).

\bibitem{Janson14}O. Janson, I. Rousochatzakis, A. A. Tsirlin, M. Belesi, A. A. Leonov, U. K. R\"o{\ss}ler, J. van den Brink, and H. Rosner, Nature Commun. {\bf 5}, 5376 (2014).


\bibitem{Romhanyi14}J. Romh\'anyi, J. van den Brink, and I. Rousochatzakis, Phys. Rev. B {\bf 90}, 140404(R) (2014).

\bibitem{Chizhikov15}V. A. Chizhikov, and V. E. Dmitrienko, J. Magn. Magn. Mater. {\bf 382}, 142 (2015).

\bibitem{Ozerov14}M. Ozerov, J. Romh\'anyi, M. Belesi, H. Berger, J.-Ph. Ansermet, J. van den Brink, J. Wosnitza, S.-A. Zvyagin, and I. Rousochatzakis, Phys. Rev. Lett. {\bf 113}, 157205 (2014).

\bibitem{Mochizuki12}M. Mochizuki, Phys. Rev. Lett. {\bf 108}, 017601 (2012).

\bibitem{Petrova11}O. Petrova, and O. Tchernyshyov, Phys. Rev. B {\bf 84}, 214433 (2011).

\bibitem{Moutafis09}C. Moutafis, S. Komineas, and J. A. C. Bland, Phys. Rev. B {\bf 79}, 224429 (2009).

\bibitem{Makhfudz12}I. Makhfudz, B. Krueger, and O. Tchernyshyov, Phys. Rev. Lett. {\bf 109}, 217201 (2012).

\bibitem{LinSZ14b}S.-Z. Lin, C. D. Batista, and A. Saxena, Phys. Rev. B {\bf 89}, 024415 (2014).

\bibitem{DaiY14}Y. Dai, H. Wang, T. Yang, W. Ren, and Z. Zhang, Sci. Rep. {\bf 4}, 6153 (2014).

\bibitem{Tatara14}G. Tatara, and H. Fukuyama, J. Phys. Soc. Jpn. {\bf 83}, 104711 (2014).

\bibitem{Onose12}Y. Onose, Y. Okamura, S. Seki, S. Ishiwata, and Y. Tokura, Phys. Rev. Lett. {\bf 109}, 037603 (2012).

\bibitem{Schwarze15} T. Schwarze, J. Waizner, M. Garst, A. Bauer, I. Stasinopoulos, H. Berger, C. Pfleiderer, D. Grundler, Nature Mater. {\bf 14}, 478 (2015).
\bibitem{Smolenski82}G. A. Smolenski and I. E. Chupis, Usp. Fiziol. Nauk {\bf 137}, 415 (1982) [Sov. Phys. Usp. {\bf 25}, 475 (1982)].

\bibitem{Pimenov06} A. Pimenov, A. A. Mukhin, V. Y. Ivanov, V. D. Travkin, A. M. Balbashov, A. Loidl, Nature Phys. {\bf 2}, 97 (2006).

\bibitem{Katsura07}H. Katsura, A. V. Balatsky, and N. Nagaosa, Phys. Rev. Lett. {\bf 98}, 027203 (2007).
\bibitem{Rikken97}G. L. J. A. Rikken, and E. Raupach, Nature {\bf 390}, 493 (1997).

\bibitem{Rikken02}G. L. J. A. Rikken, C. Strohm, and P. Wyder, Phys. Rev. Lett. {\bf 89}, 133005 (2002). 

\bibitem{Arima08}T. Arima, J. Phys. Condens. Matter {\bf 20}, 434211 (2008).

\bibitem{Kubota04}M. Kubota, T. Arima, Y. Kaneko, J. P. He, X. Z. Yu, and Y. Tokura, Phys. Rev. Lett. {\bf 92}, 137401 (2004).

\bibitem{Jung04}J. H. Jung, M. Matsubara, T. Arima, J. P. He, Y. Kaneko, and Y. Tokura, Phys. Rev. Lett. {\bf 93}, 037403 (2004).

\bibitem{Saito08a}M. Saito, K. Taniguchi, and T. Arima, J. Phys. Soc. Jpn. {\bf 77}, 013705 (2008).

\bibitem{Saito08b}M. Saito, K. Ishikawa, K. Taniguchi, and T. Arima, Phys. Rev. Lett. {\bf 101}, 117402 (2008).

\bibitem{Takahashi12}Y. Takahashi, R. Shimano, Y. Kaneko, H. Murakawa, and Y. Tokura, Nature Phys. {\bf 8},121 (2012).

\bibitem{Bordacs12}S. Bord\'acs, I. K\'ezsm\'arki, D. Szaller, L. Demko, N. Kida, H. Murakawa, Y. Onose, R. Shimano, T. R\~o\~om, U. Nagel, S. Miyahara, N. Furukawa, and Y. Tokura, Nature Phys. {\bf 8}, 734 (2012).

\bibitem{Miyahara11}S. Miyahara and N. Furukawa, J. Phys. Soc. Jpn. {\bf 80}, 073708 (2011).
\bibitem{Belesi12}M. Belesi, I. Rousochatzakis, M. Abid, U. K. R\"o{\ss}ler, H. Berger, and J.-Ph. Ansermet, Phys. Rev. B {\bf 85}, 224413 (2012).

\bibitem{Miller10}K. H. Miller, X. S. Xu., H. Berger, E. S. Knowles, D. J. Arenas, M. W. Meisel, and D. B. Tanner, Phys. Rev. B {\bf 82}, 144107 (2010).

\bibitem{Okamura13}Y. Okamura, F. Kagawa, M. Mochizuki, M. Kubota, S. Seki, S. Ishiwata, M. Kawasaki, Y. Onose, Y. Tokura, Nature Commun. {\bf 4}, 2391 (2013).

\bibitem{Okamura15}Y. Okamura, F. Kagawa, S. Seki, M. Kubota, M. Kawasaki, and Y. Tokura, Phys. Rev. Lett. {\bf 114}, 197202 (2015).

\bibitem{Mochizuki15}M. Mochizuki, Phys. Rev. Lett. {\bf 114}, 197203 (2015).
\bibitem{Gurevich96}A. G. Gurevich and G. A. Melkov, {\it Magnetization Oscillation and Waves} (CRC Pess, New York, 1996).

\bibitem{KezsmarkiCD}I. K\'ezsm\'arki, S. Bord\'acs, P. Milde, E. Neuber, L. M. Eng, J. S. White, H. M. R{\o}nnow, C. D. Dewhurst, M. Mochizuki, K. Yanai, H. Nakamura, D. Ehlers, V. Tsurkan, and A. Loidl, Nature Materials {\bf 14}, 1116 (2015).
\bibitem{Mochizuki14}M. Mochizuki, X. Z. Yu, S. Seki, N. Kanazawa, W. Koshibae, J. Zang, M. Mostovoy, Y. Tokura, and N. Nagaosa, Nature Mater. {\bf 13}, 241 (2014).

\bibitem{KongL13}L. Kong, and J. Zang, Phys. Rev. Lett. {\bf 111}, 067203 (2013).

\bibitem{LinSZ14a}S.-Z. Lin, C. D. Batista, C. Reichhardt, and A. Saxena, Phys. Rev. Lett. {\bf 112}, 187203 (2014).

\bibitem{Iwasaki14}J. Iwasaki, A. J. Beekman, and N. Nagaosa, Phys. Rev. B {\bf 89}, 064412 (2014).

\bibitem{Schutte14}C. Schutte, and M. Garst, Phys. Rev. B {\bf 90}, 094423 (2014).

\bibitem{Finazzi13}M. Finazzi, M. Savoini, A. R. Khorsand, A. Tsukamoto, A. Itoh, L. Duo, A. Kirilyuk, Th. Rasing, and M. Ezawa, Phys. Rev. Lett. {\bf 110}, 177205 (2013).

\bibitem{Koshibae14}W. Koshibae, and N. Nagaosa, Nature Commun. {\bf 5}, 5148 (2014).

\bibitem{Ogawa15} N. Ogawa, S. Seki, Y. Tokura, Sci. Rep. {\bf 5}, 9552 (2015).

\bibitem{Krawczyk14}M. Krawczyk, and D. Grundler, J. Phys.: Condens. Matter {\bf 26}, 123202 (2014).

\bibitem{MaF15}F. Ma, Y. Zhou, H. B. Braun, and W. S. Lew, Nano Lett. {\bf 15}, 4029 (2015).

\bibitem{Hirobe15}D. Hirobe, Y. Shiomi, Y. Shimada, J. Ohe, and E. Saitoh, J. Appl. Phys. {\bf 117}, 053904 (2015).

\bibitem{Shimada15}Y. Shimada, and J. Ohe, Phys. Rev. B {\bf 91}, 174437 (2015) 

\bibitem{Kovalev14}A. A. Kovalev, Phys. Rev. B {\bf 89}, 241101(R) (2014).

\bibitem{Nii14}Y. Nii, A. Kikkawa, Y. Taguchi, Y. Tokura, and Y. Iwasa, Phys. Rev. Lett. {\bf 113}, 267203 (2014).

\bibitem{LiuYH13}Y.-H. Liu, Y.-Q. Li, and J. H. Han, Phys. Rev. B {\bf 87}, 100402(R) (2013).

\bibitem{Janoschek13}M. Janoschek, M. Garst, A. Bauer, P. Krautscheid, R. Georgii, P. B\"oni, and C. Pfleiderer, Phys. Rev. B {\bf 87}, 134407 (2013).

\bibitem{Zivkovic14}I. {\v Z}ivkovi\'c, J. S. White, H. M. R{\o}nnow, K. Pr{\v s}a, and H. Berger, Phys. Rev. B {\bf 89}, 060401(R) (2014).

\bibitem{LinSZ14c}S.-Z. Lin, C. Reichhardt, C. D. Batista, A. Saxena, J. Appl. Phys. {\bf 115}, 17D109 (2014).

\bibitem{Levatic14}I. Levati\'c, V. {\v S}urija, H. Berger, and I. {\v Z}ivkovi\'c, Phys. Rev. B {\bf 90}, 224412 (2014).

\bibitem{Giess80} E. A. Giess, Science {\bf 208}, 938 (1980).

\bibitem{LinYS73}Y. S. Lin, J. Grundy, and E. A. Giess, Phys. Lett. {\bf 23}, 485 (1973).

\bibitem{YuXZ12b}X. Z. Yu, M. Mostovoy, Y. Tokunaga, W. Zhang, K. Kimoto, Y. Matsui, Y. Kaneko, N. Nagaosa, and Y. Tokura, Proc. Natl Acad. Sci. USA {\bf 109}, 8856 (2012).

\bibitem{Nagao13}M. Nagao, Y.-G. So, H. Yoshida, M. Isobe, T. Hara, K. Ishizuka, and K. Kimoto, Nature Nanotech. {\bf 8}, 325 (2013).

\bibitem{YuXZ14}X. Z. Yu, Y. Tokunaga, Y. Kaneko, W. Z. Zhang, K. Kimoto, Y. Matsui, Y. Taguchi, and Y. Tokura, Nature Commun. {\bf 5}, 4198 (2014).

\bibitem{Heinze00}S. Heinze, M. Bode, A. Kubetzka, O. Pietzsch, X. Nie, S. Bl\"ugel, and R. Wiesendanger, Science {\bf 288}, 1805 (2000).

\bibitem{Heinze11}S. Heinze, K. von Bergmann, M. Menzel, J. Brede, A. Kubetzka, R. Wiesendanger, G. Bihlmayer, and S. Bl\"ugel, Nature Phys. {\bf 7}, 713 (2011).

\bibitem{Romming13}N. Romming, C. Hanneken, M. Menzel, J. E. Bickel, B. Wolter,
K. von Bergmann, A. Kubetzka, and R. Wiesendanger, Science {\bf 341}, 636 (2013).

\end{thebibliography}
\end{document}